\documentclass[aps,prd,groupedaddress,preprint,tightenlines,eqsecnum,nofootinbib,
]{revtex4}
\usepackage{graphicx,epsf,amssymb,amsbsy,amsfonts,amssymb,amsmath,mathtools}
\usepackage[hidelinks]{hyperref}
\usepackage[usenames]{color}

\newcommand{\eq}{\begin{equation}}
\newcommand{\eqe}{\end{equation}}

\newcommand{\eqa}{\begin{eqnarray}}
\newcommand{\eqae}{\end{eqnarray}}

\newcommand\eea{\end{eqnarray}}
\newcommand\bea{\begin{eqnarray}}
 
\def\tr{\text{tr}} 

\def\R{\mathbb{R}}
\def\C{\mathbb{C}}

\def\Z{\mathbb{Z}}

\def\d{\partial}
\def\<{\langle}
\def\>{\rangle}
\def\+{\dagger}
\def\t{\theta}
\def\Th#1#2{\vartheta{\tiny\begin{bmatrix}
{#1}\\
{#2}
\end{bmatrix}}}

\begin{document}

\titlepage

\title{Temperature-reflection I: field theory, ensembles, and interactions}

\author{David A McGady \\
Niels Bohr International Academy \\
17 Blegdamsvej K\o benhavn 2100, Denmark\\
\href{mailto:mcgady@nbi.ku.dk}{mcgady@nbi.ku.dk}
\vspace{0.5cm}}

\begin{abstract}
In this paper, we revisit the claim~\cite{01T-rex0} that many partition functions are invariant under reflecting temperatures to negative values (T-reflection). The goal of this paper is to demarcate which partition functions should be invariant under T-reflection, and why. Our main claim is that finite-temperature path integrals for quantum field theories (QFTs) should be T-reflection invariant. Because multi-particle partition functions are equal to Euclidean path integrals for QFTs, we expect them to be T-reflection invariant. Single-particle partition functions though are often not invariant under T-reflection. Several exactly solvable systems are non-invariant under naive T-reflection, but are likely invariant under an extended T-reflection. We give example systems that are T-reflection invariant but are (1) non-unitary, (2) chiral, (3) interacting, (4) non-supersymmetric, or (5) non-conformal, and (6) argue that T-reflection is unrelated to time-reversal. Finally, we study the interplay between T-reflection and perturbation theory in the anharmonic harmonic oscillator in quantum mechanics and in Yang-Mills in four-dimensions. This is the first in a series of papers on temperature-reflections~\cite{01T-rex0}.
\end{abstract}

\maketitle

\tableofcontents

\section{Introduction}\label{secIntro}

In~\cite{01T-rex0}, it was noted that broad classes of partition functions, path integrals, and supersymmetric indexes are formally invariant under reflecting temperature-parameters to negative values. This claim originated from the fact that the bosonic oscillator partition function,
\begin{align}
Z_{\phi}(\beta \omega) = \sum_{n = 0}^{\infty} e^{-\beta \omega (n + 1/2)} = \frac{1}{2 \sinh(\beta \omega/2)}~,
\end{align}
is an odd function of the inverse temperature, $\beta$:
\begin{align}
Z_{\phi}(-\beta \omega) = -Z_{\phi}(+\beta \omega)~.
\end{align}
Similarly the fermionic oscillator partition function is an even function of $\beta$, and is also invariant under temperature-reflection (T-reflection):
\begin{align}
Z_{\psi}(\beta \omega) = e^{+\beta \omega/2} + e^{-\beta \omega/2} = 2 \cosh(\beta \omega/2) = Z_{\psi}(-\beta \omega) ~.
\end{align}
As argued in~\cite{01T-rex0}, this symmetry provides the basis for the claim that various quantum field theory (QFT) path integrals, superconformal indices, and Virasoro minimal model characters are invariant under temperature-reflections. The rough argument is as follows:

QFT path integrals and/or supersymmetric indices were argued to be invariant under T-reflection in~\cite{01T-rex0} in special situations, such as free theories or certain exactly solved systems, when they are expressible in terms of infinite collections of decoupled simple harmonic oscillator partition functions. In more detail,~\cite{01T-rex0} focused on QFT path integrals and supersymmetric indices that took the schematic form:
\begin{align}
Z_{\rm QFT} (\beta) = \prod_{n} Z_{\phi}(n \beta \omega) ~ \prod_{m} Z_{\psi}(m \beta \omega)~.
\end{align} 
The central question of T-reflection invariance, here, reduces down to a question of whether T-reflection invariance survives in the field theory limit of a quantum system. 

Because each individual decoupled oscillator was invariant under reflecting the sign of temperature, and because no oscillators were coupled, then one might reasonably conclude the full path integral $Z_{\rm QFT}(\beta)$ should be invariant under T-reflection:
\begin{align}
Z_{\rm QFT}(-\beta) = e^{i \gamma} Z_{\rm QFT}(+\beta)~. \label{eq1}
\end{align}
Section~\ref{secGoals} places this paper within the broad study of T-reflection. Section~\ref{secStructure} outlines the structure of the paper. In section~\ref{secContext} we briefly review similar ideas in the literature.

\subsection{Goals of this paper}\label{secGoals}

Any serious discussion of T-reflection should begin with three related lines of inquiry. First: Which systems can be invariant under T-reflection? On a related note, is T-reflection a previously unappreciated aspect of a known symmetry/property of nature? Is it new?

Second: Exactly how can T-reflection be realized in practice? For example: Does it act like a discrete $\Z_2$? Can T-reflection be obtained by some analytic continuation? Answering these questions would go a long way towards making T-reflection a consistently defined physical operation that is mathematically well-defined. This raises a related secondary question: Do we need to invent new mathematical definitions and extend our understanding of the mathematical functions that describe (simple) field theory path integrals, in order to make T-reflection well-defined? In~\cite{02T-rex2} and~\cite{03-Lfunctions} we show that T-reflection already provides new mathematical constructions and results within the theory of modular forms.

Third: What are the physical consequences that come from T-reflection? We argue that T-reflection is a relatively common, if previously unappreciated, corollary of general coordinate invariance in finite-temperature QFT. Does it therefore imply new consistency conditions on quantum field theories? Again, preliminary evidence suggests that it gives non-trivial information on the vacuum/Casimir energy of four-dimensional gauge-theories~\cite{04-QCD1, 05-QCD2, 06-QCD3, 07-QCD4}, and may suggest new consistency conditions in condensed matter systems~\cite{08-SPT-R}.

We divide our study of T-reflection in this paper and the other follow-up paper~\cite{02T-rex2} into three questions. In the follow-up paper~\cite{02T-rex2} we focus on the question: Does T-reflection hold in the detailed mathematical setting of modular forms and conformal field theories in two-dimensions? In this paper, we focus on the question: Where could one expect T-reflection to hold and why? Our results in this paper fall into two classes:

First: we show in-depth examples where T-reflection is, and is not, a symmetry of various QFT partition functions and path integrals. These examples show that T-reflection should not be a general symmetry of general systems. Rather T-reflection can only be a special property of full path integrals  for QFTs at finite temperature. 

Second: we study the interplay between T-reflection and perturbation theory. As emphasized in~\cite{01T-rex0}, every Virasoro minimal model character is invariant under T-reflection. Yet, these minimal models are strongly interacting conformal field theories. Because of this, T-reflection clearly survives as a symmetry of interacting systems. However, these minimal models are isolated solutions: They are exact solutions to a specific theory, found by using an infinitely constraining symmetry. In Ref.~\cite{01T-rex0}, we did not know whether T-reflection was a property of exactly solved quantum systems, or whether it could be seen in the rich environment of weakly coupled quantum field theories with approximate perturbative solutions.

In this paper, we present preliminary evidence suggesting T-reflection exists for perturbative corrections to the harmonic oscillator and to weakly coupled gauge theories in four-dimensions. Unfortunately, certain exactly solved quantum systems are not invariant under the naive T-reflection. However, we argue that they may very well be invariant under a generalized T-reflection operation (suggested by analytic continuation of the path integral).

\subsection{Structure of the paper}\label{secStructure}

In section~\ref{sec0} we expand on the main argument in~\cite{01T-rex0} in more explicit detail and explain some of the reasoning behind the manipulations. This highlights important and previously under-emphasized aspects of~\cite{01T-rex0}.  More importantly this discussion lays a foundation for a broader discussion of T-reflection in much more general contexts in the remainder of this paper and in Refs.~\cite{02T-rex2,08-SPT-R} and future follow-up projects.

In section~\ref{secCircle} we provide a general argument tying T-reflection invariance to encoding temperatures in QFT path integrals. The argument is as follows. Quantum field theories in $d$-dimensions are put at finite temperatures by first Wick rotating the time direction $t \to i t = t_E$ and then second periodically identifying the Euclidean time $t_E \sim t_E + n \beta$. This identification is called a thermal circle $S^1_{\beta}$, and places the QFT on the $d$-manifold
\begin{align}
{\cal M}_d = {\cal M}_{d-1} \times S^1_{\beta}~. \label{eqE2}
\end{align}
Specifically, the temperature $T$ is the inverse of the thermal circle's circumference: $T = 1/\beta$. To define the QFTs path integral, we integrate over all quantum fluctuations at all points on the space-time manifold. Here the manifold is ${\cal M}_{d-1} \times S^1_{\beta}$, and the path integral is
\begin{align}
Z(\beta) = \int {\cal D}[\phi] e^{-S_E[\phi]} \quad, \quad S_E[\phi] :=\int_{ S^1_{\beta} \times {\cal M}_{d-1} } \!\!\!\!\!\!\!\!\! d^d x~ {\cal L}[\phi(x)]~. \label{eqE3}
\end{align} 
Because we integrate over all points on the circle, the only remnant of the compactification comes from the lattice of identified points, $t_E \sim t_E + n \beta$, and the spectrum of quantized Kaluza-Klein (KK) excitations along the circle. Now, this lattice of identified points is equally well generated by the unit vector $+\beta$ or $-\beta$. 

Thus, we claim: \emph{the path integral in~\eqref{eqE3} is invariant under T-reflection because it should be invariant under this redundant coordinate reparametrization}. T-reflection is a discrete $\Z_2$ subgroup of general spacetime reparametrizations. Because of this, it is not surprising that it can constrain spacetime quantum numbers. This observation was not present in~\cite{01T-rex0}. 

It is new, and it has two important corollaries. First, T-reflection should constrain the energy spectrum in a general QFT placed at finite temperature. This explains a central observation in~\cite{01T-rex0}: demanding invariance under T-reflection uniquely fixes vacuum energy. In the forthcoming~\cite{03-Lfunctions}, we explore analog statements in the purely mathematical context of sum-rules for Borcherds products for modular forms. Second, T-reflection phases represent ambiguities in the phases of path integrals, and possible global gravitational anomalies. We explore this aspect of T-reflection further in the forthcoming Refs.~\cite{02T-rex2} and~\cite{08-SPT-R}. 

In section~\ref{secEnsembles} we emphasize that this argument works for QFT path integrals, and extends to multi-particle partition functions for a given set of interactions. This does \emph{not} extend to single-particle partition functions. We give an explicit example of this general phenomenon.

In sections~\ref{secQMpert} and~\ref{secQMexact} we study how T-reflection invariance interplays perturbation theory and whether it holds for exactly solved quantum mechanical (QM) potentials. Using the method of steepest ascent to study analytically continued path integrals, we find that perturbative corrections to the harmonic oscillator are invariant under T-reflection only if their coupling constants are also allowed to vary as $\beta$ moves to $-\beta$. In comparison to this, we highlight several exactly solved potentials that are not invariant under the naive $\beta$-reflection. 

We find unperturbed exact systems whose spectra are determined by kinematic quantization conditions (such as 2d CFTs) are invariant under simply exchanging $+\beta$ with $-\beta$. 
However, when interactions deform the spectrum in interesting ways, we must be more careful. More generally, we match $Z(\beta)$ with $e^{i \gamma}Z(-\beta)$ by simultaneously deforming the integration contour needed to define the path integral via the method of steepest ascent. In section~\ref{secQMpert} we study the harmonic and anharmonic oscillators. Here, we show that a deformation $\beta H \to e^{i \t}H(\t) \beta \to -\beta H(\pi) = \beta H$ exists and leads to an invariant Euclidean action at $\t = \pi$. Non-invariance comes from variations of the path integral measure, which we also show. Thus, when we can define such a deformation exists, we therefore prove:
\begin{align}
\!\!
Z(e^{i\t}\beta) := \int {\cal D}[x(\t)] e^{-S_E(\t)} 
\implies Z(\beta) = e^{i \gamma} Z(-\beta) ~.
\end{align}
Building on this, in section~\ref{secYM} we study consistent truncations of higher-dimensional QFTs. Specifically, we study perturbative corrections to Yang-Mills theory in four-dimensions, and an application of random matrix theory that yields ``universal'' phase diagrams for quantum systems which depend only on the order parameters of the system. We show perturbations to gauge theories behave in exactly the same manner as anharmonic perturbations to harmonic oscillators. Additionally, we point-out that the random matrix theory is invariant under T-reflection by relating it a consistent mode-truncation on the thermal circle. This universal approximation to dynamical descriptions of nature is invariant under T-reflection.

In section~\ref{secNOT} we explicitly discuss what T-reflection is \emph{not}. We show, via concrete examples derived in the previous sections, that T-reflection is not related to unitarity, parity (non)invariance, time-reversal (non)invariance, conformal invariance, supersymmetry, free theories, small perturbative interactions, or exact solvability. Further, T-reflection is unrelated to standard negative temperatures, where $\partial S/\partial E = \beta$ is less than zero. 

\subsection{T-reflection and other reflection symmetries in the literature}\label{secContext}

Finally, we comment that  T-reflection is similar to other operations that have been considered in several distinct contexts in the physics literature. In particular, Ref.~\cite{09-Ereflection} explicitly considered invariance of certain special field theories under analytically continuing $x \to i x$ as a possible symmetry to address the cosmological constant problem. Ref.~\cite{10-Ereflection2} postulated a symmetry where every state with energy $+E$ had a ghost partner with energy $-E$ as a way to address the cosmological constant problem. Refs.~\cite{11-Ereflection3,12-Ereflection4}, and to some extent~\cite{13-Ereflection5,14-Ereflection6}, consider a more modest symmetry where certain relativistic QFTs are invariant under reversing the sign of the metric. Again, this symmetry constrains the cosmological constant. 

Each of these proposed symmetries have similarities to, but differ from, T-reflection. Our perspective here is that, while these symmetries appear similar, our study of T-reflection is slightly different in spirit. We do not postulate T-reflection as a fundamental symmetry. Rather we observe that it is a symmetry of many physical systems. However, Refs.~\cite{09-Ereflection,10-Ereflection2} simply posit the existence of a symmetry solely to address the cosmological constant. Of the previous work, T-reflection seems akin to the signature-reversal symmetry~\cite{11-Ereflection3,12-Ereflection4}: each are redundant encodings of spacetime into the path integral. See also the follow-up paper~\cite{02T-rex2} and the forthcoming paper~\cite{08-SPT-R}. Our goal in this paper is to understand the reasons for T-reflection to exist, and where it exists. We claim T-reflection invariance holds broadly.  

\section{The original argument for T-reflection invariance}\label{sec0}

In this section, we revisit and expand the original argument that quantum field theory path integrals are invariant under T-reflections. We do so in the specific context of two-dimensional conformal field theories (2d CFTs), placed on the two-torus. Our focus on 2d CFTs is a matter of convenience: the two-torus is the simplest compact and factorized spacetime manifold, and 2d CFTs are subject to the infinite-dimensional Virasoro symmetry. Our main examples are the exactly solved minimal models. One of them, the Ising minimal model, plays a special role as it is dual to a theory of free fermions on the two-torus.

The structure of this section is as follows. First, in section~\ref{secHard} we discuss why T-reflection may be subtle in quantum field theories, even though it is clearly a property of decoupled harmonic oscillators. Second, in section~\ref{secOld} we repeat the original argument in Ref.~\cite{01T-rex0} that the (left-moving sector of the) free scalar and the free fermion CFT on the two-torus is invariant under T-reflection, up to an overall temperature-independent phase. Finally, in sections~\ref{secOld2} and~\ref{secMMs} we exploit the duality between the free fermion 2d CFT and the Ising model 2d CFT to show that all Virasoro minimal model characters are invariant under T-reflection. The manipulations in section~\ref{secOld} when applied to the strongly interacting minimal models in section~\ref{secMMs} then imply that T-reflection is a symmetry of both free and interacting CFTs, and does not depend in any way on parity, unitarity, or even modularity.

\subsection{T-reflection and the field theory limit}\label{secHard}

As commented in section~\ref{secIntro}, the harmonic oscillator partition function is invariant under T-reflection, up to a phase of $e^{i \gamma} = -1$. This can be understood by analytic continuation. Writing the partition function in terms of the variable $q := e^{-\beta \omega}$, we have
\begin{align}
Z(\beta) = \sum_{n = 0}^{\infty} e^{-\beta \omega(n+1/2)} = \frac{e^{-\beta \omega/2}}{1-e^{-\beta \omega}} = \frac{q^{1/2}}{1-q}~. \label{eqZ1e1}
\end{align}
There is a single isolated pole in this entire function of $q$. This pole is located at $q = 1$, and corresponds to the divergence of this the oscillator at high temperatures, where $|T| \gg \omega$. 

In this language, the positive-temperature partition function is an expansion about the point $q = 0$. The pole at $q = 1$ sets the radius of convergence of this positive-temperature expansion in the complex-$q$ plane to unity. Because there is only one pole in the complex-$q$ plane, we can continue the partition function from $|q| < 1$ to the entire $q$-plane. Note that $|q| < 1$ corresponds to complex values of $\beta$ with real positive parts: the positive-temperature region. When $|q| > 1$, we have ${\rm Re}(\beta) <0$ and thus negative temperatures. 

If we expand the oscillator partition function about $1/q = 0$, we find
\begin{align}
Z(1/q) = \frac{q^{-1/2}}{1-q^{-1}} ~,\label{eqZ1e2}
\end{align}
and thus $Z(1/q) = -Z(q)$ for the bosonic oscillator. We revisit this operation in \emph{considerable} detail in section~\ref{secQMpert}.

\begin{figure}[t] \centering
\includegraphics[width=0.95\textwidth]{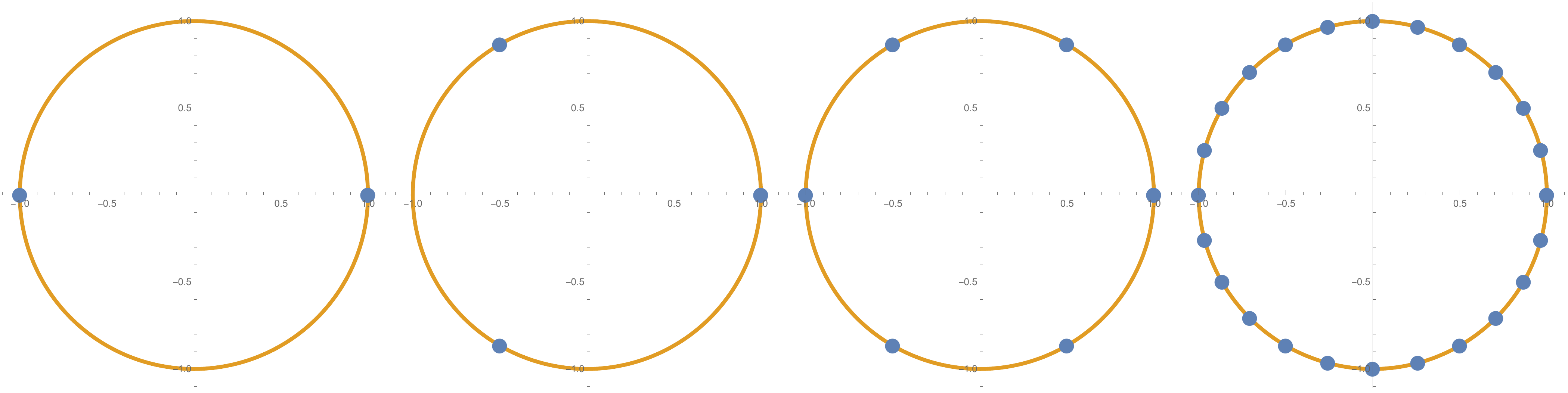} 
\caption{A plot of the location of poles of the truncated path integral for a free scalar CFT on the spacetime manifold $S^1_{L} \times S^1_{\beta}$, together with the circle $|q| = 1$. The plot is in the complex-$q$ plane, where $q = e^{-\beta/L}$ is the Boltzmann suppression factor in the free scalar CFT path integral. The infinite-temperature point point $\beta = 0$ corresponds to $q = 1$: Hence we call $|q| = 1$ the ``infinite-temperature circle''.  We depict poles in the $q$-plane for  $N =$  2, 3, 6 and 24 modes. Poles accumulate at rational angles as $N \to \infty$, ruining continuity on $|q| = 1$. This obstruction prevents naive analytic continuations within the $q$-plane that connect expansions about $q = 0$ and $1/q = 0$.} 
\label{F1_ScalarPoles}
\end{figure}

We now consider an ensemble of $N$ decoupled oscillators whose characteristic frequencies are integer-spaced: The $k^{th}$ oscillator has characteristic frequency $\omega_k = k \omega$. The partition function for this ensemble is,
\begin{align}
Z_N(q) = \prod_{k = 1}^{N} \frac{q^{k/2}}{1-q^k} = q^{\frac{1}{2} \sum_{\ell = 1}^{N} \ell^{+1}} \prod_{k = 1}^{N} \frac{1}{1-q^k}~. \label{eqZNe1}
\end{align}
Clearly, this has $N$ poles on the unit circle $q = 1$. For any finite $N$, we can analytically continue from the $|q| < 1$ ``positive-temperature'' region to the $|q| > 1$ ``negative-temperature'' region. The ensemble is invariant under T-reflection up to an overall phase:
\begin{align}
Z_N(1/q) = (-1)^{\sum_{\ell = 1}^{N} \ell^{+0}} Z_N(q)~. \label{eqZNe2}
\end{align}
In this context, the most natural explanation how and why the oscillator partition function at positive and negative temperatures are related to each other comes from analytic continuation within ``the'' complex-$q$ plane. This is depicted in figure~\ref{F1_ScalarPoles}. Nevertheless, from this perspective it is still an almost complete surprise that $Z(q) \propto Z(1/q)$. We explain this fact in sections~\ref{secCircle} and~\ref{secQMpert}.

However, this natural continuation becomes significantly more subtle for the simple free massless scalar on the two-torus. We revisit the full mathematical treatment of these subtleties in~\cite{02T-rex2}. To underscore this subtlety, note that even the simplest free conformal field theory path integrals resembles an infinite collection of decoupled bosonic oscillator partition functions. This happens for the following reason. 

General spatial manifolds have an infinite number of extended modes with well-defined wave-numbers $k_n$. Each mode $k_n$ can be occupied either a finite or an infinite number of times, depending on statistics of the quantum field. As such, the occupation numbers for each individual mode are given by an oscillator-like generating function whose characteristic frequency is given by the wave-number of the underlying mode. Generally speaking $\omega_n = \sqrt{k_n^2 + m^2}$. So if the field is bosonic, we would have
\begin{align}
Z_{n} = \sum_{\ell = 0}^{\infty} d(\ell) e^{-\beta \omega_n (\ell + 1/2) }~,
\end{align}
where $d(\ell)$ counts the degeneracy of the mode. If we set $\omega_n = n \omega$ and $d(\ell) = 1$, we recover the oscillator partition function in Eq.~\eqref{eqZ1e1}. This sum has $n$ poles on the circle $|q| = 1$. 

Because a free QFT does not couple distinct modes $(k_n,\omega_n)$, its path integral is the product over all the generating functions for each individual mode. Higher-momentum modes $\omega_n \sim n \omega$ inescapably create order $\sim \!n$ poles on the $|q| = 1$ circle. Thus, the full ensemble of modes in the full path integral has a dense set of poles on the $|q| = 1$ circle. This circle separates positive and negative temperatures in ``the'' complex $q$-plane. Consequently, this prevents analytic continuation within the obvious $q$-plane. In the following sections, we argue T-reflection invariance nevertheless survives in the field theory limit, despite the fact that we cannot analytically continue from $|q| < 1$ to $|q| > 1$ within the naive $q$-plane.

\subsection{T-reflection invariance for the free massless scalar in two dimensions}\label{secOld}

In the remainder of section~\ref{sec0}, we study conformal field theories in two dimensions (2d CFTs). We will place them on the spatial manifold $S^1_L$. Further, we compactify the temporal direction onto the thermal circle, $S^1_{\beta}$. This gives the flat two-torus, $T^2 = S^1_L \times S^1_{\beta}$.

We begin with free scalars that are periodic along each one-cycle of the two-torus. Because they are periodic along the spatial direction, the scalar excitation momenta are quantized to be $p_n = 2 n \pi/L$. As scalars are bosons, there is no limit to the number of excitations of any given mode. And because the theory is free, the energy of a collection of excitations is given by the sum of the energies of each individual excitation. Thus, the path integral is
\begin{align}
Z(q) = {\rm tr}\big[e^{-\beta H}\big] = \frac{1}{q^{1/24}} \sum_{n = 0}^{\infty} p(n) q^n  = \frac{1}{q^{1/24}} \prod_{n = 1}^{\infty} \frac{1}{1-q^n}~, \label{eqQFT1}
\end{align}
where $q^{-1/24}$ is the sum of the zero-point energies of each individual mode (Casimir energy), $p(n)$ counts the number of unrestricted partitions of $n$, and each factor of $1/(1-q^n)$ counts occupations of the $n^{th}$ excitation mode $p_n = 2 \pi n/L$. As commented in section~\ref{secHard}, while analytic continuations connect expansions of the partition function for any finite collection of $N$ modes about the point $q = 0$ to expansions of the partition function about $1/q = 0$, when we take the field theory limit and set $N = \infty$, this is no longer possible.\footnote{It is important to note that the finite product, $\prod_{\ell = 1}^{N} 1/(1-q^{\ell}) = \sum_{n = 0}^{\infty} p_N(n)$ has the interpretation as the generating function for $p_N(n)$, the partitions of an integer $n$ into at most $N$ parts. Physically, this finite product is the partition function for $N$ scalar particles. We obtain field theory by sending $N \to \infty$.}

Physically, simple QM systems and non-interacting QFTs built from these quantum systems differ only in the number of such decoupled quantum systems. We argue that this is the main difference between quantum mechanical systems and free quantum field theories constructed from a ``large'' (infinite) number of copies of these prototypical quantum systems. When the numbers of these systems are taken to be ``large", i.e. when one approaches the field theory limit, divergences appear. For instance the zero point energy of a QFT zero point energy is the sum of an infinite number of non-zero zero-point energies of each decoupled mode. 

It is important to disentangle the quantities which diverge as $N \to \infty$ from those that do not, and to recognize their significance. The argument in~\cite{01T-rex0} that T-reflection survives taking the field theory limit hinges on doing this process both at positive and negative temperatures for the finite-$N$ regularization of the field theory path integral. We now review this argument.

To begin, we consider the free energy for $N$ decoupled harmonic oscillators with frequencies $\omega_n = n/L$, each of which is separately invariant under T-reflections. For positive $\beta > 0$, the free energy of this ensemble of $N$ oscillators is 
\begin{align}
- \log Z_N(+\beta) 
&:= \sum_{n=1}^{N} \frac{\beta \omega n}{2}  + \sum_{n=1}^{N} \log(1 - e^{-\beta \omega n}) ~.
\label{fB1}
\end{align}
The two sums in Eq.~\eqref{fB1} play qualitatively distinct roles: The first sum is simply the Casimir energy of the quantum system, and the second term is the generating function for the degeneracies of any given excitation of the system. Because the oscillator frequencies increase with $n$, $\omega_n \leq \omega_{n+1}$, the sum of zero-point energies diverges as $N \to \infty$. In contrast to this, the second sum over $\log(1-e^{-\beta \omega_n})$ is absolutely convergent for fixed and positive $\beta/L$. In detail, the higher order terms are exponentially smaller than lower order terms:
\begin{align}
\log(1-e^{-\beta (n+1)/L}) \to \log(1) + e^{-\beta (n+1)/L} = e^{-\beta (n+1)/L} \ll e^{-\beta n/L} \quad {\rm for} \quad \beta/L > 0~ .
\end{align}
However, when $\beta < 0$, this second sum, which does nothing other than count the degeneracies of any given excitation of the collection of $N$ oscillators, will diverge. 

Thus before we take the field theory limit at negative temperature, we must isolate the terms in this $\beta < 0$ sum which will diverge as $N \to \infty$ from those which converge. As for the positive temperature case, the convergent terms give degeneracies of excitations above the vacuum, while the divergent terms give information about the Casimir energy and, now, the T-reflection phase that tracks the sign of $\beta$. So, we rewrite Eq.~\eqref{fB1} evaluated at $-\beta < 0$,
\begin{align}
\log Z_N(-\beta) 
&= -\sum_{n=1}^{N} \frac{(-\beta) n \omega}{2}  - \sum_{n=1}^{N} \log(1 - e^{-(-\beta) n \omega}) 
\label{fB2a} \\
&= -\sum_{n=1}^{N} \frac{(-\beta) n \omega}{2}  - \sum_{n=1}^{N} \log\bigg(e^{+\beta n \omega}(1 - e^{-\beta n \omega}) \, (-1) \bigg) \label{fB2b} \\ 
&= -\bigg\{\sum_{n=1}^{N} \frac{+\beta n \omega}{2} \bigg\} - \sum_{n=1}^{N} \log(1 - e^{-\beta n \omega}) - \bigg\{ \sum_{n=1}^{N} \log (-1) \bigg\} \label{fB2c}~.
\end{align}
In Eq.~\eqref{fB2b}, we factored the logarithmic terms such that the exponentially enhanced terms and the minus-one factors which contribute to divergences as $N \to \infty$ are isolated from the exponentially suppressed factors which  count degeneracies. In Eq.~\eqref{fB2c}, the divergent terms are grouped together such that their sum trivially yields the free energy at \emph{positive} temperature, plus an additive temperature-independent term, $\sum_{n \leq N} \log(-1)$. 

This partitioning of the free energy for $Z(-\beta)$ into the free energy of $Z(+\beta)$, plus a temperature-independent (albeit divergent) constant, exactly matches that in~\cite{01T-rex0}. The physical justification for this is that for finite-$N$, we should separate the free energy into terms that count excitations above the vacuum, which are manifestly convergent as $N \to \infty$, and terms that diverge in this limit. When we do this, we recover the same divergent sum for the Casimir energy at negative temperatures, plus a temperature-independent phase. 

Taking the final step, we send $N \to \infty$ for ${\rm Exp}[-\log Z(+\beta)]$ in Eq.~\eqref{fB1} and for ${\rm Exp}[-\log Z(-\beta)]$ in Eq.~\eqref{fB2c}. We use zeta-function regularization to extract the finite parts of the divergent sum of zero-point energies and the divergent sum of $\log(-1)$ phases:
\begin{align}
Z(+\beta) 
= q^{\frac{1}{2}\zeta(-1)} \prod_{n = 1}^{\infty} \frac{1}{1-q^n} 
&= \frac{1}{q^{1/24}} \prod_{n = 1}^{\infty} \frac{1}{1-q^n} ~,\\
Z(-\beta) 
= (-1)^{\zeta(0)} q^{\frac{1}{2}\zeta(-1)} \prod_{n = 1}^{\infty} \frac{1}{1-q^n}
&=(-1)^{-1/2} \frac{1}{q^{1/24}} \prod_{n = 1}^{\infty} \frac{1}{1-q^n}~,
\end{align}
where\footnote{Note that two T-reflections yield a phase of $(-1)^{2 \zeta(0)} = -1$. In sections~V and VI of~\cite{02T-rex2}, we relate T-reflection both to a $\pi$-rotation in a plane $\beta \to e^{i \pi} \beta$ and to the modular weight $k$. Operating twice gives a $2\pi$-rotation and also gives $(-1)^{2k}$: For modular forms  like $1/\eta(q)$ with half-integer weight, $e^{2 i \gamma_R} = -1$.} 
$q = e^{-\beta \omega} = e^{-\beta/L}$. This is the physical argument for T-reflection invariance of the free scalar 2d CFT presented in~\cite{01T-rex0}. We have done this to clearly explain why the procedure and results in~\cite{01T-rex0} should be taken seriously. We now apply it to free fermions on the two-torus.  

\subsection{Free massless fermions and the Ising minimal model in two-dimensions}\label{secOld2}

Identical logic applies to the case of the four possible sets of boundary conditions for fermions on the two-torus: periodic/anti-periodic, anti-periodic/anti-periodic and anti-periodic/periodic. Their path integrals ``characters'' are,
\begin{align}
Z_{{\rm NS-NS}}(q) 	= q^{ -\frac{1}{48}} \prod_{n = 0}^{\infty}(1+q^{n+1/2}) \quad &, \quad
Z_{{\rm NS-R}}(q) 	= q^{ -\frac{1}{48}} \prod_{n = 0}^{\infty}(1 -q^{n+1/2}) ~, \\
Z_{{\rm R-NS}}(q) 	= q^{+\frac{1}{24}} \prod_{n = 0}^{\infty}(1+q^{n}) \quad \qquad &, \quad
Z_{{\rm R-R}}(q) 	= q^{+\frac{1}{24}} \prod_{n = 0}^{\infty}(1 -q^{n}) = 0 ~.
\end{align}
Here NS corresponds to anti-periodic boundary conditions, R corresponds to periodic boundary conditions, and the first label corresponds to boundary conditions on the $S^1_{L}$ and the second label corresponds to boundary conditions on $S^1_{\beta}$. Note that the Casimir energies of each fermion character can be expressed as a regularized sum of the naive zero-point energies of each decoupled fermionic oscillator. For example,
\begin{align} 
\!
E_{\rm CAS}^{\rm NS-R} = \frac{1}{2} \sum_{n = 0}^{\infty} \frac{(-1)}{(n+1/2)^s}	\bigg|_{s = -1} = -\frac{1}{48} ~~ , ~~ 
E_{\rm CAS}^{\rm R-NS} = \frac{1}{2} \sum_{n = 0}^{\infty} \frac{(-1)}{(n+0)^s}	\bigg|_{s = -1} = +\frac{1}{24}~~. \!
\end{align}
These results come from the Hruwitz zeta-function, $\zeta_H(s,x)$. At $s = -1$ we know $\zeta_H(-1,x)$ evaluates to $x/2-x^2/2-1/12$. Each non-trivial fermion character can be recovered from a finite product, truncated to $N$ factors. Repeating the analysis of section~\ref{secOld}, we see that the positive-temperature $\log Z(q)$ and the negative-temperature $\log Z(1/q)$ again have an identical structure, up to a temperature-independent constants. For example,
\begin{align}
\log \frac{Z^{(N)}_{{\rm NS-R}}(q)}{Z^{(N)}_{{\rm NS-R}}(1/q)} 	
= \sum_{n = 0}^{N} \frac{\log( -1) }{(n+1/2)^s}\bigg|_{s = 0} \quad , \quad 
\log \frac{Z^{(N)}_{{\rm R-NS}}(q)}{Z^{(N)}_{{\rm R-NS}}(1/q)} 
= \sum_{n = 0}^{N} \frac{\log(+1) }{(n)^s}\bigg|_{s = 0} = 0~ .
\end{align}
Regularizing them in the same way that we regularize the zero-point energies, we find that the coefficient of $\log(-1)$ vanishes for $\log Z_{\rm NS-R}$; because $\log (+1) = 0$ we know that $\log Z_{\rm R-NS}$ automatically has a trivial phase. Hence, we derive that the four free fermion path integrals are totally T-reflection invariant: 
\begin{align}
Z_{{\rm NS-NS}}(q) 	= Z_{{\rm NS-NS}}(1/q) \quad &, \quad
Z_{{\rm NS-R}}(q) 	= Z_{{\rm NS-R}}(1/q)~~~~ , \label{eqFinv12} \\
Z_{{\rm R-NS}}(q) 	= Z_{{\rm R-NS}}(1/q) \quad &, \quad
Z_{{\rm R-R}}(q) 	= 0 = Z_{{\rm R-R}}(1/q) ~. \label{eqFinv34}
\end{align}
Crucially, we used the same regularization procedure for the free scalar, and isolating the convergent generating functions for energies and degeneracies of excitations above the vacuum for these regularized path integrals, in keeping with section~\ref{secOld}. However, more is true. As we shall show in the next section, this automatically implies that all interacting 2d CFT minimal model path integrals are automatically invariant under T-reflection.

\subsection{T-reflection invariance for minimal models: interactions, unitarity, and parity}\label{secMMs}

In this section we show that an infinite collection of strongly interacting 2d CFTs on the two-torus, the Virasoro minimal models, are invariant under T-reflection. These path integrals are representations of the Virasoro algebra, and are defined by a finite collection of conformal primary operators and their tower of Virasoro descendants. Descendants of a primary operator ${\cal O}_{\Delta}$ with scaling dimension $\Delta$ take the form,
\begin{align}
{\cal O}^{n_1 \cdots n_N}_{N+ \Delta} = \prod_{k = 1}^{N} (L_{-k})^{n_k} {\cal O}_{\Delta} \quad , \quad \sum_{k = 1}^{N} n_k k = N~. \label{eqMM-1}
\end{align}
Clearly, there are $p(N)$ total descendants of a conformal primary at level $N$. When all of these descendants are nontrivial and present in the spectrum, the family of conformal descendants of ${\cal O}_{\Delta}$ within a CFT of central charge $c$ is given by the generating function 
\begin{align}
\chi_{\Delta,c}(q) = q^{\Delta-c/24} \sum_{N = 0} p(N) q^n = \frac{q^{\Delta - (c-1)/24} }{\eta(q)}~, \label{eqMM0} 
\end{align}
where $\eta(q)$ is the Dedekind eta-function. 

Virasoro minimal models are CFTs whose primary operators have null descendants. Because they are null, this forces the generating function to be modified. In particular, if a CFT has both a central charge and an operator with conformal scaling dimension given by
\begin{align}
c(p,p') = 1 - 6 (p-p')^2/pp' \quad {\rm and} \quad h_{r,s}^{p,p'} = (p r - p's)^2/(4pp') - (c-1)/24~, \label{eqMM1}
\end{align}
then the primary operator has null descendants with degeneracies counted by the function
\begin{align}
\chi_{(r,s)}^{p,p'}(q) = 
  \frac{1}{\eta(q)} \sum_{n = -\infty}^{+\infty} q^{(2pp' \, n \, + (pr \mp p's))^2/4pp'} 
- \frac{1}{\eta(q)} \sum_{n = -\infty}^{+\infty} q^{(2pp' \, n \, - (pr \mp p's))^2/4pp'} \, . \, \label{eqMM2}
\end{align}
CFTs with central charges and a primary operators with conformal scaling dimensions given in Eq.~\eqref{eqMM1} are called Virasoro minimal models. They are characterized by the integers $p,p'$ and often denoted ${\cal M}(p,p')$. Minimal models are said to be strongly interacting because the conformal scaling dimensions in~\eqref{eqMM1} differ significantly from the naive conformal scaling dimension,$1$.

If a given operator has a negative conformal scaling dimension, $h<0$, then its correlation functions grow with increasing distance and the operator has negative norm descendants. Fixing all operators in a minimal model to have positive scaling dimensions fixes $p = p' + 1$. For this reason, all minimal models with $p = p' +1$ are called unitary minimal models~\cite{21-Yellow}. When $p \neq p' +1$, the minimal model is said to be non-unitary. 

Our goal in this section is to prove that all Virasoro minimal model characters are invariant under T-reflection. We do so in two steps. First, we note that all characters of the minimal model corresponding to the 2d Ising model on the square lattice at criticality are equal to (linear combinations of) free fermion characters. Thus they are automatically invariant. Second, we directly show that this same conclusion follows from the properties of the projection condition in Eq.~\eqref{eqMM2}. The manipulations leading to the characters in Eq.~\eqref{eqMM2} to be invariant for the Ising minimal model automatically carry-through for \emph{all} minimal model characters, unitary or not. In this way, we show general invariance.

\subsubsection{Fermions and the Ising minimal model}\label{secMM34}

Recall that free fermions on the two-torus are dual to the interacting 2d CFT that describes the Ising 2d model on the square lattice at criticality. The conformal scaling dimensions of the Ising model on the square lattice at criticality exactly match the three non-trivial characters of the first unitary minimal model, ${\cal M}(4,3)$. Invariance of free fermions implies invariance of this interacting theory. We show the invariance of ${\cal M}(4,3)$ in three ways:
\begin{enumerate}
\item By rewriting the Ising ${\cal M}(4,3)$ minimal model characters directly in terms of the fermion path integrals/characters, which are completely invariant under T-reflection. This method is special to the Ising model; it does not generalize. 
\item By exploiting the structure of the Ising ${\cal M}(4,3)$ minimal model characters of to rewrite them in terms of a five-fold infinite product, via the quintuple product identity, This method follows the logic in section~\ref{secOld}, and generalizes to show that all characters with $p' = 3 r$ and otherwise arbitrary $p, p'$ and $s$ are invariant under T-reflection. 
\item As pointed-out in~\cite{01T-rex0}, by writing each of the two projection operators in the definition of the character for the Ising ${\cal M}(4,3)$ minimal model in terms of a three-fold infinite product. This method implies all Virasoro minimal model characters are separately invariant under T-reflection, independent of the value of $p,p'$ or $r,s$.
\end{enumerate}
We show the first two methods, here in this subsection. We show the third in subsection~\ref{secMMgen}, below. To begin, we recall the standard identities (see for example chapter 10.3 in~\cite{21-Yellow}),
\begin{align}
&\chi_{(1,1)}^{4,3}(q) = \frac{1}{2} \bigg( Z_{NS-NS}(q) + Z_{NS-R}(q) \bigg)~,~  
\chi_{(1,3)}^{4,3}(q) = \frac{1}{2} \bigg( Z_{NS-NS}(q) - Z_{NS-R}(q) \bigg)~, \nonumber\\
&\chi_{(1,2)}^{4,3}(q) = Z_{R-NS}(q) = q^{1/24} \prod_{n = 0}^{\infty} (1+q^n)~. 
\end{align}
These identities are straightforward to show from Eqs.~\eqref{eqMM1} and~\eqref{eqMM2}. From Eqs.~\eqref{eqFinv12} and~\eqref{eqFinv34}, we see that $\chi_{i,j}(q) = \chi_{i,j}(1/q)$.

We can see this yet a different way. First, we recall the five-fold product identity:
\begin{align}
&\sum_{n \in \mathbb{Z}} Y^{\frac{3k^2+k}{2}}\left(X^{3k}-X^{-(3k+1)} \right) \\
&= \prod_{n = 1}^{\infty} (1-Y^n) (1-Y^{n-1}/X)(1-Y^n X) (1-Y^{2n-1}X^2) (1-Y^{2n-1}/X^2) \, \nonumber .
\end{align}
If it is possible to equate $X = q^{rs} = q^{p' s/3}$ and $Y = q^{2 pr} = q^{2 pp'/3}$, then the character with $p' = 3r$ can be written as a single infinite product,
\begin{align}
\chi_{(r,s)}^{p,p'}(q)\bigg|_{p' = 3r}
=& -\frac{q^{\frac{1-c}{24}+h_{r,s}+rs}}{\eta(q)} \prod_{A =1}^{\infty} 
\left(1-(q^{2pr})^{A}\right)  
\left(1-(q^{2pr})^{A-1}q^{ -rs}\right)  
\left(1-(q^{2pr})^{A}   q^{+rs}\right)  \nonumber\\
&\qquad \qquad \qquad \qquad \left(1-(q^{2pr})^{2A-1} q^{ -2rs}\right)  
\left(1-(q^{2pr})^{2A-1} q^{+2rs}\right)  \label{prodChar}
\end{align}
If we use the regularization procedure in section~\ref{secOld} and use the Hurwitz zeta function $\zeta_H(-1,x) = -(x^2-x)/2-1/12$ to regularize the divergent zero-point energies, we can show that this is invariant under T-reflection. This particular analysis shows that all characters with $p' = 3 r$ and otherwise arbitrary $p, p'$ and $s$ are invariant under T-reflection.

In addition to showing all characters of the Ising model ${\cal M}(3,4)$ and are invariant, this proof shows that every character of ${\cal M}(2,p)$ is invariant under T-reflection for any $p$. However, this direct appeal to infinite product representations is insufficient to show general minimal model characters are invariant. To proceed further, we return to the explicit evaluations of Ref.~\cite{01T-rex0}.

\subsubsection{General minimal models}\label{secMMgen}

Finally as pointed-out in~\cite{01T-rex0}, each of the two infinite sums in the projection in Eq.~\eqref{eqMM2} are separately invariant under T-reflection. To see this, note that each of the infinite sums in Eq.~\eqref{eqMM2} is a Jacobi theta-function $\vartheta_{00} (z,Q)$ with an infinite product representation:
\begin{align}
\vartheta_{00}(Q^\alpha,Q) :&= q^{\frac{(pr \mp p's))^2}{4pp'}} \sum_{n = -\infty}^{+\infty} (q^{2pp'}) ^{\frac{n^2}{2}} (q^{pr \mp p's}) ^{n} \nonumber\\
&= Q^{\frac{\alpha^2}{2}} \prod_{n = 1}^{\infty} (1-Q^n) (1+ Q^{n+\alpha-\frac{1}{2}}) (1+ Q^{n-\alpha-\frac{1}{2}}) ~. \label{Theta2}
\end{align}
Here $Q = q^{2 pp'}$, $\alpha = pr \mp p's/2pp'$, and we used Jacobi's triple-product identity for $\vartheta_{00}(Q^{\alpha},Q)$ in the last line. 

Now, T-reflection inverts $q$ and so it also inverts $Q$. Regularizing the divergent phases and divergent sums of zero-point energies with the Hurwitz zeta-function, we obtain:
\begin{align}
\label{theta_inversion}
Q^{-\frac{\alpha^2}{2}}\vartheta_{00}(Q^{-\alpha},Q^{-1}) 
=& \frac{Q^{-\frac{\alpha^2}{2}} \vartheta_{00}(Q^{+\alpha},Q^{+1}) (-1)^{\zeta(0)}}
{Q^{\zeta(-1)+\zeta(-1, \frac{1}{2}-\alpha)+\zeta(-1, \frac{1}{2}+\alpha)}} 
= (-1)^{\zeta(0)} \, Q^{+\frac{\alpha^2}{2}}\vartheta_{00}(Q^{\alpha},Q) \, .
\end{align}
The $T$-reflection symmetry of each Virasoro minimal model character then follows immediately from the fact that these projection operators are always divided by $1/\eta(q)$, which is invariant up to the same overall factor of $(-1)^{\zeta(0)}$. As a result, the phases exactly cancel, and we are left with the equation,
\begin{align}
\chi^{p,p'}_{(r,s)}(q) = \chi^{p,p'}_{(r,s)}(1/q) 
\quad,
\quad \forall {\cal M}(p,p') 
\quad {\rm and} 
\quad 1 \leq r \leq p'-1 \quad 1 \leq s \leq p-1~.
\end{align}
This proof matches the specialized techniques of sections~\ref{secOld},~\ref{secOld2}, and subsection~\ref{secMM34}. Additionally it generalizes them to every possible minimal model. Regardless of whether the model is unitary or non-unitary, and contains negative-norm states whose correlation functions that grow with separation, it is invariant under T-reflection.

Because minimal model partition functions are written as finite (sesquilinear) combinations of Virasoro minimal model characters, this implies that all Virasoro path integrals are completely invariant under T-reflection. Importantly, this invariance is independent of parity: Minimal model characters for left- or right-moving degrees of freedom are separately invariant. Under T-reflection any left-moving minimal character is mapped back into a left-moving character, namely itself. Because of this, we infer that T-reflection is decoupled from parity. For this same reason, T-reflection cannot be related to orientation-reversal in any obvious way, as orientation reversals are exchange left- and right-movers.

In conclusion, in section~\ref{sec0} we repeated the argument originally made in~\cite{01T-rex0} that all Virasoro minimal models were invariant under T-reflection. The specific manipulations in this section are very much tied to either the fact that the QFT is free, or to the fact that the QFT is a 2d CFT that tightly constrained by an infinite-dimensional Virasoro symmetry and by modularity. The argument of sections~\ref{secHard}--\ref{secOld2} extends to all free QFTs, but not to interacting ones.  This leads us to sections~\ref{secCircle} and~\ref{secQMpert}. 

In section~\ref{secCircle}, we give a general argument for the existence of T-reflection invariance. Further, a parallel of this argument for 2d CFTs yields other highly nontrivial properties of their path integrals not discussed in this paper: modularity. Making the connection between section~\ref{secCircle} and modular invariance of 2d CFTs is a main focus of the follow-up paper~\cite{02T-rex2}. 

In section~\ref{secQMpert}, we expand our results in another direction: attempting to probe whether T-reflection is a good symmetry of perturbative partition functions, or whether it can only be checked to be a property of exactly solved QFTs. To make this precise, we are forced to again consider contour deformations. These deformations play a crucial role in sections~\ref{secQMpert},~\ref{secQMexact}, and ~\ref{secYM} and in the follow-up~\cite{02T-rex2}.

\section{T-reflection and the thermal circle}\label{secCircle}

In section~\ref{sec0} we gave a simple argument that T-reflection is a symmetry of certain special field theory path integrals. In this section, we give a general argument that ties invariance under reflecting $\beta \to -\beta$ to a redundant expression for how path integrals of QFTs are put at finite temperatures. The argument is extremely straightforward. 

We put a quantum field theories in $d$-dimensions at finite temperature by first analytically continuing time to Euclidean-time, $t \to i t := t_E$, and then compactifying the Euclidean time direction by identifying $t_E$ with $t_E + n \beta$. Here, $\beta$ is the inverse temperature. Making this identification corresponds to compactifying Euclidean time onto a circle of circumference $\beta$. This circle is called the thermal circle and is denoted by $S^1_{\beta}$. Compactifying in this way leads the QFT at inverse temperature $\beta$ to be placed on the following background $d$-manifold,
\begin{align}
{\cal M}_d = {\cal M}_{d-1} \times S^1_{\beta}~. \label{eqTS1}
\end{align}
Now, recall that QFT path integrals integrate over all quantum fluctuations at every point on the spacetime manifold. Because the manifold is ${\cal M}_{d-1} \times S^1_{\beta}$, the QFT path integral is thus
\begin{align}
Z(\beta) = \int {\cal D}[\phi] e^{-S_E[\phi]} \quad, \quad S_E[\phi] :=\int_{ S^1_{\beta} \times {\cal M}_{d-1} } \!\!\!\!\!\!\!\!\! d^d x~ {\cal L}[\phi(x)]~. \label{eqTS2}
\end{align} 
In evaluating the path integral, we integrate-out the circle. The only remnant of the thermal compactification comes from the quantized Kaluza-Klein (KK) momenta along the thermal circle, which a dictated by the lattice of points, $t_E \sim t_E + n \beta$. 

Crucially, both the lattice of identified points and the corresponding spectra of KK momenta are equally well generated by the unit vector $+\beta$ as by the unit vector $-\beta$. Hence the analytic structure of the path integral has no bearing on the sign of $\beta$. For this reason, we claim that the path integral in~\eqref{eqTS1} is invariant under T-reflection: \emph{The path integral should be invariant under this redundant coordinate reparametrization $+\beta \to -\beta$}. 

Pursing this argument to its logical conclusion, we are forced to relate T-reflection invariance to a coordinate transformation that is not continuously connected to the identity. Relating T-reflection to a reparametrization invariance of spacetime suggests T-reflection is a remnant of general relativity. This exactly matches the fact that T-reflection is sensitive to vacuum energies~\cite{01T-rex0}. To see this, suppose that $Z(\beta) = {\rm tr}[e^{-\beta H}]$ is T-reflection invariant:
\begin{align}
Z(-\beta) = e^{i \gamma} Z(+\beta)~.
\end{align}
Then suppose that we shift the spectrum by an overall constant $\Delta$. The path integral of the shifted Hamiltonian, $Z_{\Delta}(\beta) = {\rm tr}[e^{-\beta(H +\Delta)}]$ is only T-reflection invariant when $\Delta = 0$:
\begin{align}
Z_{\Delta}(+\beta) &= e^{-\beta \Delta} Z(\beta) \\
\implies Z_{\Delta}(-\beta) &= e^{+\beta \Delta} Z(-\beta) = e^{+\beta \Delta} \big( e^{i\gamma} Z(+\beta) \big) = e^{i\gamma} e^{+2 \beta \Delta} Z_{\Delta}(+\beta)~.
\end{align}
If we define the vacuum energy of the system to be the lowest-energy state in the trace over the spectrum of the Hamiltonian, then we see that T-reflection invariance can only happen for one vacuum energy of a particular system. 

This striking statement is a technically obvious corollary of T-reflection invariance. Physically, it suggests that T-reflection is tied to a spacetime symmetry. This is now explained by relating T-reflection invariance to an invariance under a discrete coordinate reparametrization. Therefore it is a remnant of general relativity, capable of constraining vacuum energies.

In this light, T-reflection is a discrete $\Z_2$ subgroup of general spacetime reparametrizations. This observation was not present in~\cite{01T-rex0}. It is new, and it has two corollaries. Each corollary is potentially quite important. 

First, T-reflection should constrain the energy spectrum in a general QFT placed at finite temperature. This explains a central observation in~\cite{01T-rex0}: demanding invariance under T-reflection seemed to uniquely fix vacuum energy. In the forthcoming~\cite{03-Lfunctions}, we explore analog statements in the purely mathematical context of sum-rules for Borcherds products for modular forms. It is important to emphasize that spacetime symmetries are the only symmetries that \emph{can} be sensitive to vacuum energies. Supersymmetry, conformal invariance, or modularity are known to fix spacetime quantum numbers in special systems with a high degree of symmetry. T-reflection, however, seems to be a ubiquitous and new spacetime symmetry. Of these \emph{four} symmetries capable of fixing vacuum energies---supersymmetry, conformal invariance, modular invariance, and T-reflection invariance---T-reflection is the only one that should be present in generic field theories.

Second, T-reflection phases represent ambiguities in the phases of path integrals, and possible global gravitational anomalies. That a phase explicitly on a redundant choice of the sign of $\beta$ strongly resembles a global gravitational anomaly~\cite{15-GGA, 16-GGA2, 17-HeteroticII, 18-PolchinskiII}. As emphasized in the second paper~\cite{02T-rex2}, the logic that T-reflection invariance is due to a redundancy in how the compact spacetime is encoded in field theory path integrals naturally extends to highly nontrivial (but well-known) statements about the properties of path integrals for two-dimensional conformal field theories on the two-torus. 

Identical logic implies that 2d CFT path integrals should be invariant under the standard modular S- and T-transformations, \emph{as well as} a new so-called ``R-transformation''. This R-transformation is the direct analog of T-reflection in the context of 2d CFTs, as it reverses the sign of the temperature-parameter. In~\cite{02T-rex2} and particularly in~\cite{03-Lfunctions}, we show that this suggests new statements about the mathematical functions that describe 2d CFT path integrals known as modular forms. 

In~\cite{02T-rex2} and~\cite{08-SPT-R}, we study whether the T-reflection phase represents a genuine global gravitational anomaly phase in considerable detail. In particular in~\cite{08-SPT-R}, we focus on whether a non-trivial T-reflection phase indicates new consistency conditions on symmetry protected topological phases in insulators. This line of inquisition can be potentially very fruitful: global gravitational anomalies have been used extensively in the past to classify consistent field theories. Famously, the $SO(32)/\Z_2$ and $E_8 \times E_8$ as the unique gauge-groups for the heterotic string that are free of global gravitational anomalies related to modular transformations~\cite{17-HeteroticII}. More recently, similar ideas have been used to classify topological phases in strongly correlated many-body systems in condensed matter~\cite{19-SPT1, 20-SPT2}.

\section{Non-invariance in the canonical ensemble}\label{secEnsembles}

We have emphasized that quantum field theoretic path integrals should be invariant under T-reflection. However, we have restricted our use of the term ``partition function''. The reason for this restriction is simple and important. It is the focus of this section. 

In essence, only the multi-particle (or grand canonical) partition function has a simple relationship to the path integral in QFT: Particle creation is possible in both the grand canonical ensemble and in quantum field theory. However, particle number is fixed when one is in the canonical ensemble. On these grounds we only expect grand partition functions to exhibit symmetries of the full path integral, symmetries such as T-reflection.

Indeed, it is straightforward to construct infinite classes of examples where the single-particle ``canonical'' partition function is not invariant under T-reflection, even though the multi-particle ``grand canonical'' partition function for the same system is invariant. 

For example, the gauge invariant descendants of the gauge-field $A^{\mu}$ with conformal scaling dimension $\Delta = 1$ on the flat four-manifold $\R^{3} \times S^1$. (Gauge-fields play an important role in section~\ref{secYM}.) The single-particle excitations of this conformal primary are generated by acting on the conformal primary $A^{\mu}$ with the generators of special conformal transformations, $K_{\mu} = i \d_{\mu}$, and modding-out by the equation of motion and gauge-invariance constraints. 

This process yields the generating function for the conformal scaling dimensions and degeneracies of the gauge-invariant descendants of $A^{\mu}$~\cite{04-QCD1}:
\begin{align}
Z_{\rm CAN}(q) = 1- \frac{(1+q)(1-4q + q^2)}{(1-q)^3} = 6 q^2 + 16 q^2 + \cdots ~. \label{eqZcan?}
\end{align}
The $q$-series starts with $6 q^2$, which matches the six-components of $F^{\mu \nu}$ in four-dimensions.

A conformal transformation maps $\R^{3} \times S^1$ to $S^{3}_R \times S^1_{\beta}$. In this setting, $q$ has meaning as a Boltzmann factor, $q= {\rm Exp}[-\beta/R]$, and $Z_{\rm CAN}(q)$ in Eq.~\eqref{eqZcan?} can be directly interpreted as the canonical partition function for a free CFT coming from $A^{\mu}$. It is clear that this canonical partition function is \emph{not} invariant under T-reflection~\cite{01T-rex0}:
\begin{align}
Z_{\rm CAN}(q^{-1}) = 1- \frac{(1+q^{-1})(1-4q^{-1} + q^{-2})}{(1-q^{-1})^3}  = 1 + \frac{(1+q)(1-4q + q^2)}{(1-q)^3}  \propto \!\!\!\!\!\! \slash ~~ Z_{\rm CAN}(q)~. \label{eqZcant!}
\end{align}
We now show that, while the canonical partition for this free Maxwell theory is not invariant under $q$-inversion, the grand canonical (multi-particle) partition function is invariant up to an overall temperature-independent phase. The map between canonical partition functions and grand canonical partition functions is particularly simple for free field theories like this:
\begin{align}
Z_{\rm GC}(q) := {\rm Exp} \bigg[ -\sum_{n = 1}^{\infty} \frac{1}{n}\bigg(Z_{\rm CAN}(q^n) - Z_{\rm CAN}(0)\bigg) \bigg] ~.~\label{eqZmap}
\end{align}
From this, we can show that they corresponding grand canonical partition function for $A^{\mu}$ is
\begin{align}
Z_{\rm GC}(q) = q^{\zeta(-3) - \zeta(-1)}  \prod_{n = 2}^{\infty} \bigg(\frac{1}{1-q^n}\bigg)^{2(n^2-1)}~.
\end{align}
Here $\zeta(-3) - \zeta(-1)$ is the Casimir energy. This object is invariant under $q$-inversion. This is a specific example of a general phenomenon. Canonical partition functions are generally not invariant under $q$-inversion. However, their grand canonical partners are generally invariant.

\section{T-reflection for harmonic and anharmonic oscillators}\label{secQMpert}

In section~\ref{sec0}, showed that QFT path integrals have a dense set of poles (and zeros) on the boundary between 
positive and negative temperature regions of ``the'' complex-$\beta$ plane. This barrier prevents naive analytic continuation in the strict field theory limit. We then argued that the 2d CFTs given by free scalars, free fermions, and all  Virasoro minimal models are nonetheless T-reflection invariant. Going beyond this, in section~\ref{secCircle} we argued that generic path integrals of QFTs at finite-temperature should be invariant under T-reflection. 

Further, in section~\ref{secEnsembles} we stressed that while the argument for T-reflection invariance of path integrals naturally extends to the related multi-particle (``grand canonical'') partition functions, it does \emph{not} extend to single-particle (``canonical'') partition functions. We provided the explicit example of the partition function for a massless vector $A^{\mu}$: its single-particle partition function $z_v(q)$ is not invariant under T-reflections. However, its multi-particle partition function is invariant under T-reflection~\cite{01T-rex0,05-QCD2,06-QCD3,07-QCD4}.

Before this current section, all of our examples have been either free or exactly solved. This begs several questions: First, can T-reflection be a property of perturbative path integrals or partition functions? Second, is T-reflection invariance a symmetry of all simple and exactly solved systems? We focus on T-reflection and perturbation theory in this section and in section~\ref{secYM}. In this section, we focus on quantum mechanical (QM) systems. In section~\ref{secYM}, we focus on perturbative treatments of quantum field theories in higher-dimensions.

Our discussion of perturbative corrections to the harmonic oscillator in QM is structured as follows. First, in section~\ref{secSHO} we revisit the T-reflection invariance of the unperturbed simple harmonic oscillator. We use the method of steepest ascent to study the analytically continued path integral for the harmonic oscillator, as $\beta \to e^{i \t} \beta \to -\beta$. We show that at either end-point of this trajectory, the Euclidean action is invariant. 
As the Euclidean path action is simply $S_E = \smallint_\beta dt H$, we can interpret this as requiring $H \to H( \t) \to -H$ along the path $\beta \to e^{i \t} \beta \to -\beta$. (Note: $dt$ rotates with $\beta$.) Although highly unusual, this way of looking at T-reflection reveals why it should be a classical symmetry of the path integral: $+\beta H = (-\beta) (-H)$. In~\cite{02T-rex2} we find the $S^1_{\beta}$ zero-mode path integral measure is responsible for the T-reflection phase in $Z(-\beta) = e^{i\gamma} Z(+\beta)$, which spoils absolute T-reflection invariance.

Second, in section~\ref{secAHO} we study anharmonic corrections to the oscillator partition function. Looking at explicit examples, we see that $\beta$-reflection alone is not a symmetry of arbitrary anharmonic perturbations to the oscillator. However, we see that the perturbative corrections from an anharmonic term $\lambda_n x^n$ are invariant if we send
\begin{align}
\beta \to -\beta
\quad 
{\rm simultaneously~ with}
\quad 
\lambda_n \to i^{n+2} \lambda_n ~.
\end{align}
This transformation is entirely in keeping with the results of section~\ref{secSHO}: In order to send the full anharmonic Hamiltonian to its reverse as $\beta$ follows the path $\beta \to e^{i \t} \beta \to -\beta$, we must send $H \to H(\t) \neq H$. In section~\ref{secYM} we give evidence that exactly analogous results may directly extend and hold for perturbative corrections to gauge theories in $d$-dimensions. 

\subsection{Harmonic oscillator path integral}\label{secSHO}

T-reflection is visible at the level of the partition function of the harmonic oscillator
\begin{align}
Z(\beta) = \sum_{n = 0}^{\infty} e^{-\beta \omega (n + 1/2 + \Delta)} = \frac{e^{-\beta \omega \Delta}}{2 \sinh(\beta \omega/2)} = \frac{e^{-\beta \omega \Delta}}{\beta \omega} \prod_{k =1}^{\infty} \frac{1}{2\pi k + \beta \omega}  \frac{1}{2\pi k - \beta \omega} ~,
\end{align}
whose zero-point energy has been set to the canonical and T-reflection invariant value of $\Delta + 1/2 = 1/2$. We seek to understand this partition function as the path integral for a free scalar QFT on the one-dimensional manifold $S^1_{\beta}$. 

As a first step, we must show how to reproduce the oscillator partition function $1/2 \sinh(\beta \omega/2)$ from a Euclidean path integral on the thermal circle for ``positive'' $\beta$
\begin{align}
Z_E(\beta) = \int {\cal D}[x(t_E)] e^{-S_{E(\beta)}[x(t_E)] } = \int {\cal D}[x(t_E)] e^{-\int_0^{\beta} dt_E H[x(t_E)] }~,
\end{align}
where $t_E \sim t_E + n \beta$ for every integer $n$. To do this, we can discretize the spatial direction into $N$ intervals. This converts the Euclidean action from an integral into a discrete sum:
\begin{align}
S_{E(\beta)}[x(t)] = \int_0^{\beta} dt_E \frac{1}{2}\bigg[ \bigg(\frac{dx}{dt_E}\bigg)^2 + \omega^2 x^2 \bigg]  \longrightarrow \frac{\beta}{2 N} \sum_{n=1}^N \bigg[ \bigg(\frac{x_{n}-x_{n-1}}{\beta/N}\bigg)^2 + \omega^2 x_n^2\bigg]~.
\end{align} 
To evaluate the path integral, we Fourier-transform from position-variables $x_n$ to Fourier-modes $y_k$. This basis diagonalizes the Hamiltonian, and the path integral factorizes into:
\begin{align}
Z_E^{(N)}(\beta) &= \int d^Ny ~ {\rm Exp} \left[ - \sum_{k=0}^{N-1} y_k y_{-k} \left(4 \sin^2\left(\frac{k \pi}{N}\right) + \left(\frac{\omega \beta}{N}\right)^2\right) \right]~. 
\end{align}
Straightforward manipulation for fixed-$\beta$ yields
\begin{align}
Z_E^{(N)}(\beta)= \frac{N}{\omega \beta} \prod_{k=1}^{N-1} \left\{ 4 \sin^2\left( \frac{\pi k}{N} \right) + \left(\frac{\omega \beta}{N}\right)^2 \right\}^{-\frac{1}{2}} 
= \frac{1}{\omega \beta} \prod_{k=1}^{N-1} \left\{ 1 + \frac{(\omega \beta)^2}{4 N^2 \sin^2\left( \frac{\pi k}{N} \right)} \right\}^{-\frac{1}{2}}~.\!
\end{align}
It is important to emphasize that each of the $N$ factors in this product correspond to the Gaussian integral for the $k^{th}$ Fourier mode on the thermal circle. So, when we send $N \to \infty$, this finite product over $N$ Fourier modes changes to an infinite product over the infinite set of winding modes around the thermal circle:
\begin{align}
\lim_{N \to \infty} Z_E^{(N)}(\beta) = \frac{1}{\omega \beta} \prod_{k = 1}^{\infty} \left\{ 1 + \frac{(\omega \beta)^2}{(2 \pi k)^2} \right\}^{-\frac{1}{2}} \left\{ 1 + \frac{(\omega \beta)^2}{(-2 \pi k)^2} \right\}^{-\frac{1}{2}} 
= \prod_{k \in \Z} \frac{1}{2 \pi i k - \beta \omega} ~.
\end{align}
As claimed, this expression is manifestly equal to $1/2\sinh(\beta \omega/2)$. This proves a direct equality between the partition function for the simple harmonic oscillator and the Euclidean path integral for a free scalar on the thermal circle. 

Having established this equality, we now revisit why this is invariant under T-reflection. The first answer is extremely straightforward: As discussed in section~\ref{secCircle}, the only remnant of the thermal circle in finite-temperature path integrals can come from the quantization conditions on the Kaluza-Klein (KK) modes. Quantization comes only from the periodicity condition $t_E \sim t_E + n \beta$. This is equally well described by $+\beta$ or by $-\beta$. Exchanging $+\beta$ and $-\beta$ does not change the spectrum of KK modes, and it so can not change the path integral. This explains why the original simple harmonic oscillator is invariant under T-reflections.

This answer however only allows us to view T-reflection as a discrete $\Z_2$ operation. In particular, it does not involve analytic continuation. Analytic continuation is far more constraining than a discrete $\Z_2$-reflection. As such, we now attempt to show that T-reflection can be realized as an analytic continuation of the quantum oscillator as $\beta \to e^{i \t} \beta \to -\beta$.

\subsubsection{Lefshetz thimble}\label{secThimble}

To do so, we have to explicitly describe how to define the path integral $Z(\t)$ once its parameters, such as $\beta$, are deformed away from their real values into the complex plane. To begin, we first fix the branch-choice of these integrals at $\beta = {\rm Re}(\beta) > 0$. Second, we study how the integrals change as we deform $\beta$ to $e^{i \t} \beta$. Because $\beta$ has been deformed, the Hamiltonian density in the Euclidean path integral is also deformed,
\begin{align}
S_E[x(t)] = \beta \int_{0}^{1} dt_E~ H(t_E)[x(0)] \to S_E(\t) =  e^{i \t} \beta \int_{0}^{1} dt_E ~H(t_E)[x(\t)]~.
\end{align}
The final path integral is defined to be an integral over the quantum field (here, $x[t]$) along the thermal circle $S^1_{\beta}$. 

When the deformation is trivial, this is just the original path integral:
\begin{align}
Z(0) = \int {\cal D}[x(t_E)] ~{\rm Exp} \big( - S_E[x(t)] \big)
\end{align}
This path integral is preformed over all real values of $x$. However, for non-zero $\t$, the Hamiltonian density and the ``volume'' of the $\t$-deformed thermal-circle change. A path integral at non-zero $\t$ is defined only once the contour of integration is also defined. 

We denote this contour $\Gamma(\t)$. It is defined to be the path of steepest ascent, where the action increases maximally as the path goes towards its boundaries. If we parametrize positions along $\Gamma(\t)$ by a real parameter $x(\t,t)$ for $t \in \R$, then $\Gamma(\t)$ must solve 
\begin{align}
\frac{\partial x(\t,s)}{\partial s} = -\overline{\bigg(\frac{\partial I(\t)}{\partial x(\t,s)}\bigg)}~,
\end{align}
where $\overline{w}$ denotes complex conjugation. Generally, there may be multiple solutions to this extremizing condition. We select the solution that decays at infinity. Thus, we require 
\begin{align}
\lim_{s \to - \infty} x(\t,s) = 0 \, .
\end{align}
This extremization procedure, where the contour of integration of a deformed path integral is required to follow the path of steepest ascent is known as the Lefshetz thimble~\cite{22-ThimbleRef1, 23-ThimbleRef2, 24-ThimbleRef3} and is the general tool used to define continuations of path integrals. (See for example~\cite{25-ThimbleRef4, 26-ThimbleRef5, 27-ThimbleRef6}).

It is particularly simple to evaluate the paths of steepest ascent for the factorized Gaussian integrals over the KK Fourier modes that define the oscillator path integral. To do so, we recall that the original mode with KK momentum $k = 2 \pi n/\beta$ has a Gaussian path integral
\begin{align}
\!\! I_{k}(\beta):=\int dy_k ~ {\rm Exp} \bigg[ y_k y_{-k} \bigg(4 \sin^2\bigg(\frac{k \pi}{N}\bigg) + \bigg(\frac{\omega \beta}{N}\bigg)^2\bigg) \bigg]~.
\end{align}
Note that the integration variable is $y_k$, the amplitude for this $k^{th}$ mode to be occupied. 

This integral has a structural difference for $k \neq 0$ and $k = 0$: when $k = 0$, the $\sin^2(\pi k/N)$ kinetic-term in the action vanishes. At the original $\t = 0$ point, we have
\begin{align}
I_{k\neq 0}(\beta) = \bigg(1 + \frac{(\omega \beta)^2}{4 N^2 \sin^2(\pi k/N )} \bigg)^{-\frac{1}{2}} \quad , \quad 
I_{k = 0}(\beta) =  \sqrt{\frac{1}{\beta^2 \omega^2}} := +\frac{1}{\beta \omega}~~ .\!\!
\end{align}
We may now define the $\t$-deformed contributions from the $k^{th}$ KK mode. 

First, we fix the branch-choice of these integrals at $\beta = {\rm Re}(\beta) > 0$. Second, we study how the integrals change as we deform $\beta$ to $e^{i \t} \beta$. We use the method of steepest ascent. To see precisely how this works, we focus on $I_{k = 0}(e^{i \t} \beta)$. By definition, it is given by
\begin{align}
I_{k =0}(e^{i \t} \beta) = \int_{\Gamma(\t)} dy_0(\t) ~ {\rm Exp}\big[- \big( \beta \omega e^{i\t} ~ y_0(\t) \big)^2\big] ~.
\end{align}
To find the integration contour $\Gamma(\t)$, we simply maximize the growth of the ``action'' in the exponential. To do so, we note that if $y_0(\t) = e^{-i \t} y_0(\t = 0)$ then as $|y_0| \to \infty$, the action has maximal real part. This is the path of steepest ascent. Thus,
\begin{align}
I_{k =0}(e^{i \t} \beta) 
&= \int_{y_0(\t = 0) \in R} d(~e^{-i \t} y_0(\t = 0)~) ~ {\rm Exp}\big[- \big( \beta \omega e^{i\t})^2 ~ (e^{-i \t} y_0(\t = 0) \big)^2\big] \\
&= e^{-i \t} I_{k = 0}(e^{i 0} \beta) = e^{- i \t} \sqrt{\frac{1}{\beta^2 \omega^2}} ~. \label{eqKK0}
\end{align}
Similar results hold for the other KK winding modes. 

If we look at $k \neq 0$, we see that the two terms in the integrand rotate by a relative phase of $e^{-2 i \t}$. Crucially, Eq.~\eqref{eqHrotFull} in the next section exactly reproduces this $e^{-2 i \t}$ counter-rotation.

\subsubsection{Creation and annihilation operators}\label{secRaising1}

In this way, we can see that the oscillator path integral $Z(\beta)$ can be defined at $\beta > 0$ and analytically continued to the same function, with a relative minus-sign when $\beta \to e^{i \t} \beta \to -\beta < 0$. In this subsection, we define how the Hamiltonian and associated creation and annihilation operators transform as $\beta$ is analytically continued to $e^{i \t} \beta$. Because the path integral is invariant as we continue from $+\beta \to -\beta$, we expect that at the end of the path $+H$ continues to $-H$. Changing the sign of $H$ indicates that the creation and annihilation operators should exchange roles. Finally, the KK zero-mode sector of this continued Hamiltonian should match the above result in Eq.~\eqref{eqKK0}. The Hamiltonian and its associated creation an annihilation operators that we define here exactly reproduces all of these expectations. To show this, we first make an ansatz for how the creation and annihilation operators behave, as a function of the deformation parameter $\t$. We then construct the Hamiltonian and its eigenvectors as a function of $\t$. Finally, we check that these results are consistent with the path integral result for the KK zero-mode in subsection~\ref{secThimble}.

We begin by defining the rotated creation and annihilation operators:
\begin{align}
\left\{ 
\begin{matrix}
a(0) ~:= a ~ \\
a^{\+}(0) := a^{\+}
\end{matrix}
\right.
\to 
\left\{ 
\begin{matrix}
a(\t) ~:=~+ \cos(\t/2) a - i \sin(\t/2) a^{\+} \\
a^{\+}(\t) := -i\sin(\t/2) a ~+ \cos(\t/2) a^{\+}
\end{matrix}
\right. ~.
\end{align}
It is straightforward to show that $[a(\t),a^{\+}(\t)] = 1$ holds, and is independent of $\t$. With these, we can define the natural extension of the Hamiltonian for general $\t$:
\begin{align}
H(0) = \frac{1}{2} \big( a a^{\+} + a^{\+} a \big) \to H(\t) := \frac{1}{2} \big(a(\t) a^{\+}(\t) + a^{\+}(\t) a(\t) \big) = a(\t) a^{\+}(\t) + \frac{1}{2}~.
\end{align}
It is important to emphasize that for $\t \neq 0$ or $\t \neq \pi$, while the Hamiltonian has a positive-definite spectrum, it is \emph{not} Hermitian. Because $(a(\t))^{\+} = a^{\+}(-\t)$, we can to show that
\begin{align}
\big(H(\t)\big)^{\+} ~= H(-\t) ~.
\end{align}
Although it is not Hermitian, $H(\t)$ has exactly the same spectra as $H(0)$. To see this, we explicitly construct the eigenstates of $H(\t)$ by making use of the unchanged commutation relation $[a(\t), a^{\+}(\t)] = 1$. We define the ground state to be the state annihilated by $a(\t)$:
\begin{align}
a(\t) |0(\t) \> = 0~. 
\end{align}
We then define excited states $|N(\t)\>$ by,
\begin{align}
|N(\t)\> := \frac{1}{\sqrt{N!}} (a^{\+}(\t))^N |0(\t)\>~.
\end{align}
Repeated use of the canonical commutation relation then implies,
\begin{align}
H(\t) |N(\t)\> = (N + 1/2) |N(\t)\>~,
\end{align}
which proves that $H(\t)$ has a positive-definite spectrum (resembling PT-symmetry~\cite{28-PT1,29-PT2,30-PT3,31-PT4}).

Following this, it is important to understand the precise relationship between $H(\t)$ and the original Hamiltonian, $H(0)$. To this end,  it is useful to note that position and momentum extend to $x(\t)$ and $p(\t)$ in the following way:
\begin{align}
\left\{ 
\begin{matrix}
x = \frac{1}{\sqrt{2}} (a^{\+}+a) \\
p = \frac{i}{\sqrt{2}} (a^{\+}-a)
\end{matrix}
\right.
\to 
\left\{ 
\begin{matrix}
x(\t) ~:= \frac{1}{\sqrt{2}} (a^{\+}(\t)+a(\t)) = e^{-i \t/2} x(0) = e^{-i \t/2} x \\
p(\t) ~:= \frac{i}{\sqrt{2}} (a^{\+}(\t)-a(\t)) = e^{+i \t/2} p(0) = e^{+i \t/2} p
\end{matrix}
\right. ~.
\end{align}
From this, it is straightforward to show,
\begin{align}
H(\t) = \frac{1}{2} + a^{\+}(\t) a(\t) = \frac{1}{2}\big( x(\t)^2 + p(\t)^2 \big) = \frac{1}{2} \big( e^{-i \t} x^2 + e^{+i \t} p^2 \big) \label{eqHrotFull}
\end{align}
where again $x = x(\t = 0)$ and $p = p(\t = 0)$ are the position and momentum operators of the original Hamiltonian. 

Now, note that the KK zero-mode on $S^1_{\beta}$ has a vanishing kinetic energy, and so $p = 0$. Thus, its entire contribution to the path integral comes form the $x(\t)^2$-term of the rotated Hamiltonian. Thus, we find the $p = 0$-sector of the path integral has the Hamiltonian,
\begin{align}
{\rm KK~zero~mode}~:~H_{p = 0}(\t) = e^{-i \t} \frac{x^2}{2} = e^{-i \t} H_{p = 0}(0)~.
\end{align}
This directly agrees with the KK zero-mode's contribution to the path-integral in Eq.~\eqref{eqKK0}. 

Now, recall that the path integral is invariant under T-reflection. This means that at the level of the classical Euclidean action, the end-points of the evolution $\beta \to e^{i \theta} \beta$ and $H \to H(\theta)$ should be identical. From this, we expect that $e^{i \t} H(\t) \to e^{i \pi} \beta H(\pi) = +\beta H(0)$. It is not difficult to show that this is indeed the case for this extended definition of $H(\t)$:
\begin{align}
\!\!
\beta H = \frac{\beta}{2}\big( x^2 + p^2) \to 
(e^{i \t} \beta) H(\t) = \frac{\beta}{2}\big( x^2 +e^{+i(2 \t)} p^2) \to 
(e^{i \pi} \beta) H(\pi) =+ \frac{\beta}{2} \big( x^2 + p^2 \big)~.\!
\end{align}
To summarize, this $\theta$-deformed Hamiltonian shares the spectrum of the original Hamiltonian. It is Hermitian only at the beginning- and end-points of the path.\footnote{Intriguingly, under the definition of the PT-transformation in~\cite{28-PT1} when acting on complex $x = e^{-i \t}|x|$ and $p = e^{i \t}|p|$, 
\begin{align}
P:
\left(
\begin{matrix}
x \\
p
\end{matrix}
\right)
\to
\left(
\begin{matrix}
-x \\
-p
\end{matrix}
\right) 
~~ , ~~
T:
\left(
\begin{matrix}
x \\
p
\end{matrix}
\right)
\to
\left(
\begin{matrix}
\overline{x} \\
-\overline{p}
\end{matrix}
\right) ~~
\implies PT\big( H(\t)  \big) = H(-\t)~,
\end{align}
we see that PT transformations on this non-Hermitian Hamiltonian $H(\t)$ acts in a way that is identical to Hermitian conjugation. It would be extremely illuminating to pursue the relationships between the now well-known and well-studied PT-symmetric non-symmetric  Hamiltonians for quantum systems that nonetheless have real spectra~\cite{28-PT1, 29-PT2, 30-PT3, 31-PT4, 32-PT5}. }

Further, the energy of the KK zero-mode of the $\t$-deformed Hamiltonian exactly matches the action for the KK zero-mode in the $\t$-deformed path integral, as dictated by the Lefshetz thimble~\cite{22-ThimbleRef1, 23-ThimbleRef2, 24-ThimbleRef3, 25-ThimbleRef4, 26-ThimbleRef5, 27-ThimbleRef6}. In addition to this, the value of $e^{i \theta} \beta H(\theta)$ behaves exactly as a classical symmetry of the path integral should behave. Namely, at either end of the deformation, it matches the original value: $(\beta) (H) = (-\beta)(-H)$. 

Finally, we note that at the end-point of the $\t$-deformation, the original creation and annihilation operators exchange roles: 
\begin{align}
\left(
\begin{matrix}
a\\
a^{\+} \!
\end{matrix}
\right)
:=
\left(
\begin{matrix}
a^{~}(\t = 0)\\
a^{\+}(\t = 0)
\end{matrix}
\right)
\to
\left(
\begin{matrix}
a^{~}(\t)\\
a^{\+}(\t)\end{matrix}
\right)
\to
\left(
\begin{matrix}
a^{~}(\t = \pi)\\
a^{\+}(\t = \pi)\end{matrix}
\right)
=
\left(
\begin{matrix}
-i a^{\+}(\t = 0)\\
-i a^{~}(\t = 0)
\end{matrix}
\right)~. \label{eqAdA}
\end{align}
This means that $x \to -i x$ and $p \to +i p$.\footnote{It would also be amusing to think of these phases as Berry phases. For example, $\t : 0 \to 2 \pi$ maps $a \to -a$ and $a^{\+} \to - a^{\+}$, and thus $\<N(0)|N(2\pi)\>$ equals $(-1)^N$. We thank Shunji Matsuura for pointing this out.} These transformation rules are distinct from, but resemble, the transformations in~\cite{09-Ereflection}. They will play a crucial role in understanding the behavior anharmonic corrections to the harmonic oscillator under T-reflection, in section~\ref{secAHO}. 

\subsubsection{Oscillators and an outer automorphism of the Virasoro algebra}\label{secRaising2}

This behavior of the creation and annihilation operators under the T-reflection contour above has an interesting connection to an algebraic argument for T-reflection for 2d CFTs first offered in~\cite{01T-rex0}. For the free scalar 2d CFT, we can rewrite the Virasoro operators in terms of individual creation and annihilation operators for given modes on the torus. Specifically,
\begin{align}
L_n = \frac{1}{2} \sum_{m \in \Z} a_{n-m} a_{m}~, \label{eqVir1}
\end{align}
where $a_{-|k|}$ is the creation operator for a mode with momentum $k$ and $a_{+|k|}$ is the corresponding annihilation operator.

In~\cite{01T-rex0}, we suggested that T-reflection invariance of 2d CFT path integrals could be explained by two facts. First, that the path integrals are representations of the Virasoro algebra. Second, the defining commutation relation for the Virasoro algebra,
\begin{align}
[L_m ~,~ L_n] = (m-n) L_{m+n} + \frac{c}{12} m(m^2-1) \delta_{m+n,0} \label{eqVir2}
\end{align}
is invariant under the outer automorphism $L_m \to -L_{-m}$. Indeed, this outer automorphism is straightforwardly related to inner-automorphisms of compact Lie algebras, which exchange raising and lowering operators an transformation that similarly fixes the lowest (highest) weight state in a given finite-dimensional representation of the compact Lie algebra~\cite{01T-rex0}. Such operations resemble T-reflection, which superficially reflects the notion of positive energies.

Now, the mapping of creation and annihilation operators under the T-reflection contour defined in subsection~\ref{secRaising1} naturally extends to the above $a_{\pm|k|}$ operators. Namely, we exchange creation and annihilation operators and multiply them by $-i$ as in Eq.~\eqref{eqAdA}:
\begin{align}
T~:~a_{+k} \to -i a_{-k} \implies T~:~L_{n} \to -L_{-n}~ . \label{eqVir3}
\end{align}
In words, the same transformation that is responsible for T-reflection invariance in the simple harmonic oscillator is also responsible for the outer automorphism of the Virasoro algebra that was conjectured in~\cite{01T-rex0} to be connected to T-reflection invariance of 2d CFTs. 

It is important to stress that we do not expect outer automorphisms to explain T-reflection invariance in all of its appearances. Spectrum generating algebras are special properties of special quantum field theories. We only mention the connection between our continuation here and the outer automorphism of the Virasoro algebra, in order to provide further evidence that the continuation throughout section~\ref{secSHO} is meaningful and ties into previous statements about T-reflection in QFT.

\subsection{The anharmonic oscillator}\label{secAHO}

We now seek to understand the interplay between T-reflection and anharmonic corrections to the oscillator potential. Just as the harmonic oscillator is the prototypical example for free quantum field theories, anharmonic perturbations are the prototypical example for perturbative quantum field theory. In section~\ref{secYM}, we show strong evidence that the same results gleaned in this section continue to apply in perturbative QFT.

The behavior of $a(\t)$ and $a^{\+}(\t)$ under the $\theta$-deformation of the harmonic oscillator play a crucial role in our understanding of T-reflection (non)invariance of perturbative corrections to the free oscillator partition function. In the free system, as we deformed $\beta$ via the path $e^{i\theta} \beta \to -\beta$, the free Hamiltonian was deformed as well: $H \to H(\t)$, which evaluates to $-H$ at the end-point of the deformation. Hence,
\begin{align}
\beta H_{\rm HO} \to (e^{i \theta} \beta) ~ (H_{\rm HO} (\theta)) \to (-\beta) ~ (- H_{\rm HO}) = +\beta H_{\rm HO}~. \label{eqAHO1}
\end{align}
where $H_{\rm HO}$ is the Hamiltonian for the exactly harmonic oscillator. Such invariance is needed and expected in order for the path integral to be even hope to be invariant (up to a phase) under T-reflection. Anharmonic perturbations alter the Hamiltonian:
\begin{align}
\beta H_{\rm HO} \to \beta H_{\rm AHO} = \beta (H_{\rm HO} + V) = \beta H_{\rm HO} + \beta ~\sum_{n = 3}^{\infty} \frac{\lambda_n}{n!} x^n~. \label{eqAHO2}
\end{align}
For this new Hamiltonian to be invariant under T-reflection, which sends $\beta \to -\beta$, we expect to that the same operation sends $V \to -V$.

Though the anharmonic perturbations ruin integrability, they are necessarily small. So the definition of $a(\t)$ and $a^{\+}(\t)$ as given in section~\ref{secSHO} remains valid when the coupling constants $\lambda_n$ are small compared to other scales. To allow for interactions to modify this picture, we also promote the $\lambda_n$ to be functions of the deformation. Because $x \to x(\t) \to -i x$ along the path $\beta \to e^{i \t} \beta \to -\beta$, individual terms in the perturbation transform as
\begin{align}
\lambda_n x^n \to \lambda_n(\t) x(\t)^n \to \lambda(\pi) x(\pi) = (-i)^n \lambda_n(\pi) x^n~. \label{eqAHO3}
\end{align}
Now, we only expect T-reflection invariance if the $\t$-deformation flips the sign of $V$ when $\t \to \pi$. Explicitly, we expect $V(\theta) \to V(\pi) = -V$. T-reflection should fail to be a property of the perturbative corrections, unless $\lambda_n(\theta)$ transforms in the following way:
\begin{align}
\lambda_n \to \lambda_n(\t) \to \lambda(\pi) = (-i)^{n+2} \lambda_n~. \label{eqAHO4}
\end{align} 
If $\lambda(\theta) x(\theta)^n$ transforms in this way at $\theta = \pi$, then $V(\pi) = -V$ for generic values of anharmonic perturbations, $\lambda_n(0)$.
Papers on PT-symmetry~\cite{31-PT4} note that Hamiltonians such as $p^2 + x^2 + (i x)^3$ have real spectra. Amusingly, T-reflection generates exactly these interactions.

Importantly, this general expectation is exactly matched both by explicit examples and by \emph{general} expressions for the perturbative corrections to the oscillator partition function from~\cite{33-AHO}. This is starkly visible in concrete examples. For example, Eqs.~(2.35) and~(2.36) of~\cite{33-AHO} for the first and second order corrections to the oscillator partition function for $H_{\rm HO} + V_{3,4}$ with $V_{3,4} = \lambda_3 x^2/3! + \lambda_4 x^4/4!$ reads,
\begin{align}
&Z(\beta) = \frac{q^{1/2}}{1-q} \bigg( 
1 + 
\frac{\lambda_4}{8} \bigg\{ (\beta \omega) \frac{(1+q)^2}{(1-q)^2} \bigg\} + 
\bigg(\frac{\lambda_3}{3!}\bigg)^2 \bigg\{(\beta \omega) \frac{11 + 38 q + 11 q^2}{(1-q)^3} \bigg\} \label{eqNaya1}\\
&+\bigg(\frac{\lambda_4}{4!}\bigg)^2 \bigg\{ 6 (\beta \omega) \frac{(1+q)(7 + 20 q + 7 q^2)}{(1-q)^3} + \frac{9}{2} (\beta \omega)^2 \frac{1 + 20 q +54 q^2 + 20 q^3 + q^4}{(1-q)^4} 
+ {\cal O}(V^3)
\bigg\}
\bigg) 
~,\label{eqAHO5}
\nonumber
\end{align}
where $q = e^{-\beta \omega}$ and $q^{1/2}/(1-q)$ is the partition function for the purely harmonic oscillator. Clearly, there are two distinct ways to make the perturbative expansion in Eq.~\eqref{eqNaya1} invariant under T-reflection, which changes the sign of $\beta$:
\begin{enumerate}
\item Send $\beta \to -\beta$ and $\omega \to -\omega$. Thus $q \to q$, while $\lambda_n$ remains untouched.
\item Send $\beta \to -\beta$ and $\lambda_n \to (-i)^{n+2} \lambda_n$. Thus $q \to 1/q$, while $\omega$ remains untouched.
\end{enumerate} 
The former option is not consistent with T-reflection, as described in section~\ref{secSHO}: here, T-reflection changed the sign of $\beta$ but left $\omega$ untouched. Further, T-reflection corresponds to inverting $q$ to $1/q$. The former option is not T-reflection as discussed in this paper.

However, the latter option is completely consistent with the general formalism developed in this section and in section~\ref{secSHO}. Indeed, it is straightforward to prove that the generating function for the perturbative corrections to the anharmonic oscillator partition function developed in~\cite{33-AHO} is completely invariant under
\begin{align} 
\beta \to -\beta 
\quad {\rm and} \quad
\lambda_n \to (-i)^{n+2}\lambda_n 
\quad . 
\end{align}
In section~\ref{secYM}, we present evidence that similar results hold in perturbative corrections to weakly coupled gauge theories in general dimensions. On the strength of the results in this section, we may conclude that T-reflection invariance requires continuing the coupling constants in a weakly-coupled theory. In this context, T-reflection seems akin to a duality between theories than to a symmetry of a given theory. See Refs.~\cite{12-Ereflection4, 34-Ereflection8} for similar remarks in a different context and in string theory.

\section{The failure of naive T-reflection for three exact models}\label{secQMexact}

In section~\ref{secQMpert} we studied in considerable detail how the harmonic oscillator, and its anharmonic perturbations, behave under T-reflection. In this analysis, we attempted to understand T-reflection as an analytic continuation from $\beta$ to $-\beta$, via the complex deformation $\beta \to e^{i \t} \beta$. Computing the path integral along this trajectory we see that it is invariant, up to a complex phase associated with the contribution from the Kaluza-Klein zero-mode. We then constructed an adiabatically $\t$-deformed Hamiltonian, $H(\t)$, and its associated $\t$-deformed creation and annihilation operators. 

We then studied whether T-reflection survives in perturbative corrections to the harmonic oscillator partition function. We studied this both via explicit examples from the literature~\cite{33-AHO} and the results for the $\t$-deformed creation and annihilation operators. We found that T-reflection is not a property of perturbative corrections to the anharmonic oscillator if the T-reflections only change the sign of $\beta \to -\beta$. Invariance is restored if we generalize the T-reflection operation so that it can modify perturbative coupling constants.

Importantly, we were able to understand invariance of the harmonic oscillator under T-reflection by studying how the $\t$-deformed Hamiltonian behaved at the end-point of the deformation. For the harmonic oscillator, we showed that $H(\t) \to H(\pi) = -H$, where $H$ is the Hamiltonian for the original harmonic oscillator. In this way, we have
\begin{align}
\!\!
\beta H \to (e^{i \t} \beta) ~ H(\t) \to (-\beta) ~(-H) = +\beta H \implies {\tr}[e^{-\beta H}] \to {\tr}[e^{-\beta e^{i\t} H(\t)}] \to {\rm tr}[e^{-\beta H}]~. \!
\end{align}
To implement this for the anharmonic oscillator we saw that we must also change the coupling constant that multiplies the interaction term $x^n$  from $\lambda_n$ to the new value $(-i)^n \lambda_n$. 

This may suggest that T-reflection invariance as depicted in section~\ref{secCircle} may need to be reinterpreted or modified in some crucial aspect. One possible option/reinterpretation is that it is broken by perturbative interactions. Another option is to interpret T-reflection as a sort of duality. In this guise, if a potential is invariant under T-reflection without modification, then it resembles a self-duality. Similarly, if T-reflection forces a change in the interactions at $-\beta$, then it may be a duality between one (or several) potentials. As commented above, Refs.~\cite{12-Ereflection4, 34-Ereflection8} make similar assertions for a related reversal in the context of string theory.

In this section, we study three exactly solved systems whose partition functions are not invariant under T-reflection in its most naive application, where we send $\beta \to -\beta$ within a partition function while leaving all other parameters unchanged. In section~\ref{secSpSp}, we study the partition function for a quantum mechanical system of two spin-1/2 states, with a spin-spin coupling. In section~\ref{secPTV}, we study the partition function for the exactly solved Poschl-Teller potential. Finally, in section~\ref{secIsing} we study Wannier's exact solution to the Ising model on the triangular lattice~\cite{36-IsingTriangle}. Each system fails to be invariant under naive $\beta$-reflection for a clear reason, a reason that easily fits into the narrative of this paper.

\subsection{Spin-spin interactions}\label{secSpSp}

Simple harmonic oscillators play a crucial role in elucidating the structure of T-reflection. Two-level systems, which we may call fermionic oscillators, have an obvious symmetry point: $E + \Delta \in \{-\omega/2, +\omega/2\}$. For $\Delta = 0$, these two-level systems are invariant under T-reflection.

If we consider a pair of spin-1/2 states that transform in the fundamental representation of $SU(2)$ and are governed via the Hamiltonian,
\begin{align}
\hat{H}:= 
\mu \big( \vec{B} \cdot \vec{S}_1 \big) + 
\mu \big( \vec{B} \cdot \vec{S}_2 \big) + 
4 \Delta \big( \vec{S}_1 \cdot \vec{S}_2 \big)~,
\label{eqSpSp1}
\end{align} 
then the eigenstates are the usual spin-singlet $|0,0\>$ and spin-triplet $|1,m\>$. These states have energies,
\begin{align}
E(|0,  0\>) =  - 3 \Delta \quad &, \quad
E(|1,  0\>) ~= ~ \Delta \\
E(|1,+1\>) =  \mu + \Delta  \quad &, \quad
E(|1, -1\>) =  \Delta  -\mu 
\label{eqSpSp2}
\end{align}
The partition function for this system is clearly not invariant under T-reflection,
\begin{align}
Z(\beta|\Delta ,\mu) = e^{3 \Delta \beta} + e^{-\Delta \beta} (1+ e^{-\mu \beta} + e^{+\mu \beta}) = Z(-\beta|-\Delta,\mu)~, \label{eqSpSp3}
\end{align}
unless we also change the sign of the spin-spin coupling $\Delta$. Note that $Z(\beta|\Delta,\mu)$ does not depend on the sign of $\mu$. This is related to the fact that compact Lie algebras have an inner automorphism that exchanges highest and lowest weight states within finite-dimensional irreducible representations. These automorphisms are discussed in section~\ref{secRaising2} and Ref.~\cite{01T-rex0}.

Fundamentally, the spin-spin interaction is electromagnetic in origin. In perturbation theory, the energy scale of the perturbation $\Delta$ is via an ${\cal O}(\alpha)$-effect. Changing the sign of $\Delta$ corresponds to changing the sign of $\alpha \to -\alpha$, which corresponds to changing $e \to \pm i e$. As we shall show in section~\ref{secYM}, this is a natural outgrowth of the analysis in section~\ref{secQMpert} when applied to perturbative corrections to gauge theory path integrals.\footnote{We will discuss this $\alpha \to -\alpha$ reflection in further detail in section~\ref{secYM}. However, it is worth noting that some exactly solved zero-temperature partition functions are invariant under coupling constant reversals, For instance, there is a famous example where the path integral for ${\cal N} = 4$ SYM~\cite{35-VafaWitten} can be exactly solved. It is written terms of the following function of the complexified gauge coupling constant $\tau := \t/2 \pi + i 4\pi/g_{\rm YM}^2$,
\begin{align}
Z(\tau) := \frac{1728}{E_4(\tau)^3 - E_6(\tau)^2}~.\label{eqVW0a}
\end{align}
Here the functions $E_4(\tau)$ and $E_6(\tau)$ are absolutely convergent sums over a lattice of points $m + n \tau$ for every non-zero integer pair $m,n$ in $\Z^2$:
\begin{align}
E_{2k}(\tau):= \frac{1}{2 \zeta(2k)} \sum_{(m,n) \in \Z^2 \setminus (0,0)} \frac{1}{(m+n\tau)^{2k}} = E_{2k}(-\tau)~.~\label{eqVW0b}
\end{align} 
Thus, we can see  functions are manifestly invariant under reflecting $\tau \to -\tau$. If we set the Yang-Mills $\t$-term to zero, $\tau = i/\alpha_{\rm YM}$. Thus, $\alpha_{\rm YM}$-reflection is an exact symmetry of these exact partition functions. We stress the equality $E_{2k}(-\tau) = E_{2k}(+\tau)$ is far from standard. It is, in a real sense, \emph{the} central observation and \emph{the} central focus of the follow-up work~\cite{02T-rex2}. We revisit $\alpha$-reversals in section~\ref{secYM} and particularly in section~\ref{secYM1}.}

Demanding T-reflection invariance for this system forces us to change the sign of the spin-spin interaction. Unlike the anharmonic oscillator, this system has a manifestly finite-dimensional Hilbert space, and hence can be thought of being under much better control. We have just shown that the ${\cal O}(\alpha)$ coupling $\Delta \vec{S}_1 \cdot \vec{S}_2$ between these separately invariant systems must change sign to preserve T-reflection.

\subsection{The Poschl-Teller Potential}\label{secPTV}


Again, the T-reflection non-invariance of spin-spin interactions seems as if it is remedied by a change in the potential. The situation, however, is more complicated for a non-relativistic particle interacting with the exactly solved Poschl-Teller potential~\cite{38-Private2},
\begin{align}
H:= -\frac{1}{2m} \frac{d^2}{dx^2} - \frac{J(J+1)}{2 m L^2}\frac{1}{ \cosh^2(x/L)}~.
\end{align}
When $J$ is a positive integer, then this potential is exactly solvable. By going to the dimensionless variable $\tilde{x} := x/L$, the Hamiltonian differential equation reduces to,
\begin{align}
\tilde{H} \psi_E(\tilde{x})= - \left(\frac{d^2}{d\tilde{x}^2} + \frac{J(J+1)}{ \cosh^2(\tilde{x})} \right) \psi_E(\tilde{x}) = (2mL^2)E  \psi(\tilde{x})~.
\end{align}
There are $J$ bound states to this potential and a continuum above the potential. Further, the energy-scale for bound-state energies is given by the width of the potential, $L$. 

The partition function for this quantum system is set by the energies and degeneracies and densities of state of the discrete bound-states and of the continuum states. They are,
\begin{align}
{\rm Bound}~&:~E_{\ell} = -\frac{\ell^2}{2mL^2} 
~~ , ~~ \forall ~\ell \in \{ 1, \cdots, J \} = \Z_J \\
{\rm Continuum}~&:~E_k = +\frac{k^2}{2 m L^2}  
~~ , ~~ \forall ~k \in \R 
~~ {\rm and} ~~ 
\rho(k)dk = \frac{1}{\pi}\bigg( 1- \sum_{\ell = 1}^{J} \frac{2\ell}{\ell^2+k^2}\bigg)dk~~. 
\end{align}
It is straightforward to show that the partition function for this system is,
\begin{align}
Z(\beta) 
&= \sum_{\ell = 1}^{J} e^{-\beta E_{\ell}} 
+ \int_0^{\infty} dk ~ \rho(k) ~ e^{-\frac{\beta k^2}{2mL^2}} \\
&= \sqrt{\frac{m L^2}{\pi \beta}} - \sum_{\ell = 1}^{J} e^{\frac{\beta}{4 mL^2} \ell^2}~ {\rm Erfc}\bigg(\sqrt{\frac{\beta}{4 mL^2} \ell^2}\bigg)
~ ,~ {\rm with} ~~ 
{\rm Erfc}(z):= \frac{2}{\sqrt{\pi}} \int_z^{\infty} dt~e^{-t^2}.
\end{align}
This partition function is not invariant under simply changing the sign of $\beta \to -\beta$. For example, expanding this for small-$\beta$ yields a sum of the form:
\begin{align}
Z(\beta) &= \frac{1}{\sqrt{-2 \pi \beta E_1}} 
+ \sum_{\ell = 1}^{J}\bigg(\sum_{n = 0}^{\infty}\frac{C_n}{2^{n/2}} (-E_{\ell}\beta)^{\frac{n}{2}}\bigg)~,
\end{align}
where $C_n$ are calculable constants. Changing $\beta \to -\beta$ causes these terms to alternate. Thus the Poschl-Teller potential is not invariant under the most naive version of T-reflection. 

However, as shown in section~\ref{secQMpert} this is not necessarily the only way to preform T-reflection. Although the simple harmonic oscillator is invariant under this naive T-reflection operation, we were only able to understand it fully after recasting T-reflection as an analytic continuation where $\beta \to e^{i \t} \beta$. At either end of this trajectory, we keep the quantity $\beta E$-fixed: we send $\beta E \to e^{i \t} \beta E(\t) \to (-\beta)(-E) = +\beta E$. It was almost an accident that T-reflection worked as a naive $\beta \to -\beta$ reflection for the harmonic oscillator, without any sophisticated analysis. (This accident is explained by the fact that the spectrum of the harmonic oscillator is solely dictated by the symmetry structure of the spacetime manifold $S^1_{\beta}$, and hence automatically encodes the $\beta \to -\beta$ redundancy discussed in section~\ref{secCircle}.)

It seems quite likely that the full and correct T-reflection operation is rather to hold $\beta/2mL^2$-fixed, while allowing $\beta$ to vary: $\beta \to e^{i \theta \beta}$. This prescription is a natural corollary of the Lefshetz-thimble/path of steepest ascent approach put forward in section~\ref{secSHO}. If we were to adopt this prescription, then it is reasonable to expect that the path integral for the Poschl-Teller potential should be totally invariant, up to overall phase factors coming from variations of the measure of the path integral. 

Indeed, it is reasonable to speculate that the naive T-reflection phase of the first term, $e^{i \pi/2} = i$, is due to the fact that sending $E_k \to -E_k$ requires sending $k \to i k$. Precisely because this is a non-relativistic system, the phase-space integral needed to define the path integral is fundamentally one over momenta, $\smallint dk$. Thus, the transformation $k \to i k$ should exactly account for this phase. 

In appendix~\ref{secRandom} we show that the non-relativistic quantum partition function for an infinite square-well is invariant under naive T-reflection, up to a phase of $e^{i \pi/2} = (-1)^{1/2}$. After this, we show the partition function for a particle on a ring threaded by an arbitrary magnetic flux is invariant under T-reflection, with a phase of $e^{i \gamma} = (-1)^{1/2}$. The conclusions in appendix~\ref{secRandom} are supported by the logic of Ref.~\cite{01T-rex0}, and by $q$-inversions of infinite product expansions of partition functions in section~\ref{sec0}, and by more sophisticated analysis in~\cite{02T-rex2} that relates T-reflection phases of a path integral to its modular weight.

\subsection{The 2d Ising model on the square and triangular lattice}\label{secIsing}

Finally, we should comment that T-reflection is \emph{not} a symmetry of certain exactly solved statistical mechanical models. For instance, Wannier's exact solution of the antiferromagnetic Ising model (with zero external field) on the 2d triangular lattice~\cite{36-IsingTriangle} is not invariant under T-reflection. This can be traced to the simple fact that the spectrum is not symmetric about zero. This, though, is not a surprise. We do not expect T-reflection to hold for generic statistical mechanical models, unless they can be derived from a full quantum field theoretic path integral on a manifold that includes the thermal circle.

\section{T-reflection in perturbative quantum field theories}\label{secYM}

In this section, we study the T-reflection invariance of consistent approximations to fully interacting QFTs. However, before moving to discuss this, we pause to recapitulate our results on both perturbative corrections to quantum systems, and on exactly solved quantum systems. In section~\ref{secQMexact}, we found two quantum mechanical potentials whose partition functions are not invariant under naive T-reflections: spin-spin interactions in section~\ref{secSpSp}, and the Poschl-Teller potential in section~\ref{secPTV}. 

Of the two potentials, the Poschl-Teller potential (and other exactly solved quantum systems) represents the most serious challenge to the evidence in section~\ref{secCircle} that T-reflection is ubiquitous. We presented an argument that suggests the Poschl-Teller potential may be invariant under T-reflection, as obtained by analytically continuing the path integral from $+\beta$ to $-\beta$ by contour deformation via the path of steepest ascent~\cite{22-ThimbleRef1, 23-ThimbleRef2, 24-ThimbleRef3, 25-ThimbleRef4, 26-ThimbleRef5, 27-ThimbleRef6}. But this argument is far from explicit. 

In its current form, the exact partition function for the Poschl-Teller potential is not invariant under changing the sign of $\beta$. It seems likely that analytic continuation is \emph{the} correct way to think about T-reflection\footnote{See the following paper~\cite{02T-rex2} for a prescription for a picture for how write T-reflection for non-trivial 2d CFTs as an analytic continuation.}, and therefore that the non-invariance of the Poschl-Teller partition function is a red-herring. But, again, this is far from certain.

However, incorporating the the spin-spin interaction of section~\ref{secSpSp} within relativistic quantum electrodynamics (QED) points the way towards ``curing'' its T-reflection non-invariance. Fundamentally, spin-spin dipole interactions are mediated by the electromagnetic field. The correction to energy levels of a given system are perturbative order-$\alpha$ effects. 

And so we seek to understand T-reflection of this potential in terms of perturbation theory in QED. In section~\ref{secQMpert} we found that T-reflection invariance of the harmonic oscillator is spoiled by general anharmonic corrections, unless they also are modified by the T-reflection operation. In sections~\ref{secYM0} and~\ref{secYM1}, we will extend the analysis of sections~\ref{secCircle},~\ref{secEnsembles}, and~\ref{secQMpert} to perturbative corrections to gauge theory path integrals. Additionally, in section~\ref{secYM1} we will provide strong evidence that supports this extension, in the detailed setting of Yang-Mills theory with maximal (${\cal N} = 4$) supersymmetry. 

Finally, in section~\ref{secRMT} we show that consistent truncations of the full tower of thermal circle Kaluza-Klein (KK) excitations within random matrix models are invariant under T-reflection. These truncated random matrix models encode the ``universal'' phase diagrams for quantum systems which depend only on the order parameters of the system. They are completely invariant under T-reflection, as their action depends on an even quadratic function temperature that comes from the kinetic energy of the KK modes on $S^1_{\beta}$.

\subsection{Free $SU(N)$ gluons coupled to adjoint matter on $S^3 \times S^1$}\label{secYM0}

Our main examples in this section and the next will be a tractable limit of $SU(N)$ gauge theory, first outlined in~\cite{39-QCD0} and then refined in~\cite{04-QCD1}. The tractable limit is the following. 

First, we consider a four-dimensional $SU(N)$ gauge-theory with gluons, $N_s$ massless scalars and $N_f$ massless  fermions. The fermions and scalars must transform in the adjoint representation of the $SU(N)$ gauge-group. This restriction is mainly made to make the state-counting, which comes later in the construction, straightforward. (As pointed-out in the refinement~\cite{04-QCD1}, similar analysis can be done for fundamental matter.)

Second, we place the gauge-theory on the compact manifold $S^3 \times S^1$. Because the spacetime manifold is compact, then the excitations in the gauge-theory are necessarily color-singlets. This happens because of the (color-)Gauss law: any excitation with non-trivial $SU(N)$-charge will emit color flux-lines. However, because the space is compact, the flux-lines can not go off to spatial infinity. Hence, they must close back on themselves. This is a purely geometrical way to achieve confinement. Crucially, it is unrelated to dynamics.

Third, we send $N \to \infty$. Because the theory is on the compact manifold $S^3 \times S^1$, its excitations are color-neutral single- and multi-trace states. The traces contain operator insertions that correspond to excitations of the vector, fermion, and scalar fields in the theory. At finite-$N$, there are relationships between single- and multi-trace states. By taking the $N \to \infty$ limit at the outset, we explicitly forbid non-trivial relationships between single- and multi-trace states.

Fourth, we fix the radius of $S^3$ the spatial manifold, $R$, to be much smaller than the strong-coupling scale of the $SU(N)$ gauge-theory: $R \Lambda_{QCD} \ll 1$. This forces the characteristic length-scale of the interactions to be in the asymptotically-free regime, and thus suppresses anomalous dimensions. Because of this, the spectrum of excitations in this theory is that of a (constrained) free theory.

Fifth, finally we fix the temperature of the theory to be less than the de-confinement temperature. In detail, we require $\beta \Lambda_{QCD} \gg 1$. Placing the gauge theory on the manifold $S^3_R \times S^1_{\beta}$ with $R \Lambda_{QCD} \ll 1 \ll \beta \Lambda_{QCD}$ forces the theory to arbitrarily weakly coupled and yet still in its confining phase.

It is not difficult to see that in this limit, the single-trace partition function is,
\begin{align}
Z_{\rm ST} = -\sum_{k = 1}^{\infty} \frac{\varphi(k)}{k} \log \left( 1-z_v(q^k) - N_s z_s(q) +(-1)^{\tilde{B} k} N_f z_f(q^k) \right)~, \label{eqYM0a}
\end{align}
where $\tilde{B}$ dictates the boundary conditions for the fermions on the thermal circle, $q = {\rm Exp}[-\beta/R]$ is the Boltzmann factor for excitations in this space, $z_v(q)$, $z_s(q)$, and $z_f(q)$ simply count the gauge-invariant descendants of conformal primaries with $(\Delta,J_L,J_R)$ given by $v: (1,\tfrac{1}{2},\tfrac{1}{2})$, $s: (1,0,0)$, and $f: \{ (\tfrac{3}{2},\tfrac{1}{2},0), (\tfrac{3}{2},0,\tfrac{1}{2})\}$. 

This single-trace partition function is the analog of the canonical-partition function: it cannot match the QCD path integral, as it is an artificial truncation to a single-particle sector of the full spectrum of excitations of the theory. It is not invariant under T-reflection, as can be seen from the structure of $z_v(q)$, $z_s(q)$, and $z_f(q)$~\cite{04-QCD1}:
\begin{align}
z_v(q) = 1 + \frac{(1+q)(1-4q + q^2)}{(1-q)^3} \quad , \quad
z_s(q) = \frac{q(1+q)}{(1-q)^3} \quad {\rm and} \quad
z_f(q) = 4\frac{q^{3/2}}{(1-q)^3} \quad. \label{eqYM0b}
\end{align}
Clearly, $1-z_v(1/q) = -(1-z_v(q))$ and $z_X(1/q) = -z_X(q)$ for $X = s, f$.\footnote{Note that $z_v(q)$ is also not invariant by itself. This non-invariance is exactly the example in section~\ref{secEnsembles} where we highlighted the fact that canonical partition functions cannot be invariant under T-reflection, even though grand-canonical partition functions may be.} From Eqs.~\eqref{eqYM0a} and~\eqref{eqYM0b}, we immediately see that the single-trace partition function is not invariant under T-reflection: $\log (1-z_v(1/q) \neq \log (1-z_v(q)$. 

Because single-trace states do not couple at large $N$, the single-trace partition function uniquely specifies the full multi-trace partition function. Using the mapping in Eq.~\eqref{eqZmap}, we obtain~\cite{04-QCD1, 39-QCD0} the full multi-trace partition function for this free theory:
\begin{align}
Z_{\rm MT}(q) = \prod_{n = 1}^{\infty} \frac{1}{1-z_v(q^n) - N_s z_s(q^n) +(-1)^{\tilde{B} n} N_f z_f(q^n)}~.\label{eqYM0c}
\end{align}
This multi-trace partition function is the generating function for the energy and degeneracy for an any excitation in the theory. It is proportional to the path integral. 

Further, it is clear that each of the factors in the product are invariant under $q \to 1/q$, up to an overall phase of $(-1)^3$. Preforming identical analysis to section~\ref{sec0} reveals
\begin{align}
Z_{\rm MT}(1/q) = (-1)^{3 \zeta(0)} Z_{\rm MT}(q) = (-1)^{-3/2} Z_{\rm MT}(q)~. \label{eqYM0d}
\end{align}
This invariance was originally noted in Ref.~\cite{05-QCD2}. It is immediately obvious that because of the T-reflection properties of $1-z_v(q) + \cdots$ in the individual factors, that if the denominator vanishes at $q = q^{\star}$, then if \emph{also} vanishes at $q = 1/q^{\star}$. As noted in~\cite{05-QCD2}, this immediately implies that the multi-trace partition function is a finite product of Jacobi theta-functions and Dedekind eta-functions.

It is instructive and useful to give two concrete examples of this phenomenon. First, for pure Yang-Mills with $N_s = N_f = 0$, we have the exact expression:
\begin{align}
Z_{\rm YM}(\tau) 
&= \prod_{n = 1}^{\infty} \frac{(1-q^n)^3}{(1+q^n)(1-q^n/(2+\sqrt{3})(1-q^n (2+\sqrt{3})} \label{eqYM0e}
\end{align}
The infinite products are exactly those of Jacobi theta functions and Dedekind eta-functions. Explicitly,  if we define $b := \cos^{-1}(2)/2\pi$, then in the notation of appendix A of~\cite{40-QCD5} this is
\begin{align}
Z_{\rm YM}(\tau) 
&= 2 \sqrt{2} i \sin\left(\pi b\right) e^{-i\pi b} \eta(\tau)^3 \frac{\eta(\tau)}{\Th{1/2}{b+1/2}(\tau)} \left[\frac{\eta(\tau)}{\Th{1/2}{0}(\tau)}\right]^{1/2} ~. \label{eqYM0f}
\end{align}
(See also Ref.~\cite{41-QCD6}.) When $N_s = 6$ and $N_f = 4$, then we have maximally supersymmetric Yang-Mills (${\cal N} = 4$ SYM). If we define $\tilde{b} = \cos^{-1}(2)/2\pi$~\cite{40-QCD5}, it is simple to show~\cite{40-QCD5}: 
\begin{align}
Z_{{\cal N} = 4}(\tau) &=\frac{1}{\eta(\tau)}
\left(\frac{\eta(\tau)}{ \Th{0}{0}(\tau)}\right)^2
\frac{ 2\cos(\pi \tilde{b}) e^{- i \pi \tilde{b} }\,\, \eta(\tau)^2}{ \Th{1/2}{\tilde{b}}(\tau) \,\, \Th{0}{\tilde{b}+\frac{1}{2}}(\tau)} ~. \label{eqYM0g}
\end{align}

The straightforward nature of this analysis seems to run counter to the argument of the preceding section, where we argued that seeing T-reflection in a full path integral is a subtle process. It begs the question: Why were we able to see T-reflection in such a simple manner? The answer to this question is as straightforward as it is fundamental: 

The partition functions in this section are the partition functions for \emph{free} theories. We have seen throughout this paper, mainly in sections~\ref{sec0} and~\ref{secQMpert} and appendix~\ref{secRandom}, that free theories are straightforwardly invariant under T-reflection. Fundamentally, the reason for this comes down to the fact that free theories are basically free decoupled oscillators. In turn, harmonic oscillators are T-reflection invariant at the very start.  As we shall show, subtleties arise when we consider the T-reflection properties of interacting theories.

\subsubsection{Immediate applications of T-reflection}

Before moving on, we briefly remind the reader that T-reflection invariance here has four immediate and powerful implications. First, as noted in Ref.~\cite{05-QCD2} and Refs.~\cite{40-QCD5, 41-QCD6} T-reflection implies that these \emph{non-supersymmetric four-dimensional gauge theories} are described in terms of functions that describe path integrals for two-dimensional conformal field theories. (Compare the above partition functions to those in section~\ref{sec0}.) This seems to imply that these four-dimensional large-$N$ gauge-theories may be dual to two-dimensional conformal field theories. This is very different than many of the current 2d-4d mappings for twisted indices for supersymmetric theories. Our 2d-4d duality is for the \emph{full} path integral of a \emph{non-supersymmetric} gauge theory in four dimensions! 

Second, intimately related to this 2d-4d duality, we see that the path integral has a modular symmetry. This modular symmetry naturally mandates that the true vacuum energy must vanish: Changing the vacuum energy would spoil modularity!

Third, as noted in Ref.~\cite{01T-rex0} and Refs.~\cite{06-QCD3, 40-QCD5}, T-reflection invariance implies that the vacuum energy is constrained to \emph{vanish} for arbitrary $N_s$ and arbitrary $N_f$. Interestingly, there is evidence that vanishing persists even beyond the conformal point~\cite{07-QCD4}, where modularity is lost.

Fourth, as noted in Ref.~\cite{05-QCD2} for QCD with fermions, there is an \emph{explicit} cancellation between bosons with energy $n$ and fermions with energy $n + 1/2$ in the twisted path integral. This happens so long as $N_s \leq 2(N_f-1)$. As emphasized in~\cite{05-QCD2}, this is an emergent boson-fermion level matching even in non-supersymmetric theories. It is an example of a much broader phenomenon of a fermionic symmetry that occurs in models of non-supersymmetric large-$N$ adjoint QCD that exhibit both volume independence and Hagedorn growth~\cite{42-QCD7}.

\subsection{One-loop corrections to supersymmetric Yang-Mills on $S^3 \times S^1$}\label{secYM1}

In this section, we will discuss T-reflection for perturbative corrections to gauge-theory path integrals. Our focus will be on four-dimensional gauge-theories, but the approach is quite general: we expect it to apply to gauge theories in $d$-dimensions. Our main discussion naturally splits into three parts. First, in subsection~\ref{secYM1a} we apply the prescription for T-reflection for the anharmonic oscillator to the anharmonic self-interactions of gauge-fields and charged matter. Second, in subsection~\ref{secYM1b} we show that applying the procedure in section~\ref{secYM1a} to the spin-spin Hamiltonian in section~\ref{secSpSp} cures the naive non-invariance of this system. Third, in subsection~\ref{secYM1c} we study concrete examples of perturbative (and non-perturbative) path integrals for limits of ${\cal N} = 4$ SYM and of non-supersymmetric Yang-Mills theory. We show strong evidence that the prescriptions for the anharmonic oscillator exactly match the explicit and concrete results. 

\subsubsection{Importing T-reflection for the anharmonic oscillator to perturbative gauge-theory}\label{secYM1a}

In section~\ref{secSHO}, we showed that the definition for the path integral for the harmonic oscillator requires $H \to H(\t) \to H(\pi) = -H(0)$ as we rotate $\beta \to e^{i \t} \beta \to -\beta$. This sends $x \to i x$ and $p \to -i p$. Because  the physics of perturbative QFT is that of very weakly coupled harmonic oscillator modes, it is natural to insert this behavior of the weakly anharmonic oscillator as $\beta \to -\beta$ into QFT. So it is natural to conjecture that path integrals for perturbative QFTs are invariant under the (extended) T-reflections
\begin{align}
\phi(x^{\mu}) \to  i \phi(x^{\mu}) \quad , \quad
\psi(x^{\mu}) \to i \psi(x^{\mu}) \quad , \quad
A^{\nu}(x^{\mu}) \to i A^{\nu}(x^{\mu}) \quad , \label{eqYM2a1}
\end{align}
where $\phi$ is a scalar field, $\psi$ is a fermion field, and $A^{\mu}$ is a massless vector-field.

This expectation comes from section~\ref{secAHO}. There, we showed that in order for the anharmonic oscillator Hamiltonian to change sign at the end of this deformation, we needed to send the $x^3$-coupling $\lambda_3$ to $-i \lambda_3$. This was borne-out by the explicit behavior of explicit the perturbative corrections to the anharmonic oscillator from~\cite{33-AHO}. So, again, it is natural to expect to preserve T-reflection in the presence of perturbative coupling, we should send
\begin{align}
g_{\rm YM} \to -i g_{\rm YM} \implies
\frac{g_{\rm YM}^2}{4 \pi} \to - \frac{g_{\rm YM}^2}{4 \pi}~. \label{eqYM2a2}
\end{align}
Clearly this reflects the sign of the fine-structure constant. It is simple to show that the transformations in Eqs.~\eqref{eqYM2a1} and~\eqref{eqYM2a2} reverse the sign of the Lagrangian-density. For example, the kinetic term for fermions minimally coupled to a non-abelian vector field and the field-strength tensor for the vector field transform as
\begin{align}
\label{eqYM2a3} 
\begin{split}
	\overline{\psi} \gamma_{\mu} (i \d^{\mu} + i g_{\rm YM} A^{\mu}) \psi 
	& \to 
- \overline{\psi} \gamma_{\mu} (i \d^{\mu} + i g_{\rm YM} A^{\mu}) \psi~, \\
F_{\mu \nu} = (\d_{\mu} A_{\nu} - \d_{\nu} A_{\mu}) + i g_{\rm YM} [A_{\mu}, A_{\nu}] 
	& \to 
i (\d_{\mu} A_{\nu} - \d_{\nu} A_{\mu}) + i^2 g_{\rm YM} [A_{\mu}, A_{\nu}] = i F_{\mu \nu}~.
\end{split}
\end{align}
This reverses the sign of the Lagrangian density ${\cal L}$:
\begin{align}
\implies 
{\cal L} 
= \overline{\psi} \gamma_{\mu} (i \d^{\mu} + i g_{\rm YM} A^{\mu}) \psi - \frac{1}{4} F_{\mu \nu}F^{\mu \nu} 
\to -{\cal L} ~. \label{eqYM2a4}
\end{align}
Surprising though this may appear, reflecting the sign of the fine-structure constant is superficially very similar to reflecting $\beta \to -\beta$ within a thermal partition function. Reflecting the fine-structure constant changes the sign of the Lagrangian in the path integral; reflecting the sign of $\beta$, when one continues in the manner defined in section~\ref{secQMpert}, changes the sign of the Hamiltonian in the thermal partition function (or path integral). We claim that $\beta \to -\beta$ and $g \to - i g$ are invariances of perturbative corrections to gauge theory path integrals.

In the remainder of this section, we present detailed evidence that this operation seems to also be a symmetry of perturbative corrections to gauge-theories.

\subsubsection{Corrections to the structure of Hydrogen and to ${\cal N} = 4$ SYM as $\alpha \to -\alpha$}\label{secYM1b}

Taking the prescription of sections~\ref{secAHO} and~\ref{secYM1a} seriously, we should only expect perturbative corrections to finite-temperature gauge-theory path integrals to be T-reflection invariant if we send
\begin{align}
\beta \to -\beta \quad {\rm and} \quad g \to -i g ~. \label{eqYM2b3}
\end{align}
Here, $g$ is the cubic coupling between free plane-wave states and $\beta$ is the inverse temperature.
 
Now, as observed in Ref.~\cite{43-QCD8}, the lattice gauge-theory of a twisted $SU(N)$ gauge-theory is invariant under reversing the sign of $g_{\rm YM}^2$. In this example, as $g^2$ is reflected to $-g^2$, the action around plaquettes changes sign. This is a direct analog of what we observed in the anharmonic oscillator. Crucially, the path integral in~\cite{43-QCD8} is invariant under this reversal.\footnote{Because this is a discretized and finitization of the theory, it may be possible to connect the invariance for this discretized gauge-theory on the hypercubic lattice to the T-reflection invariance for Onsager's solution to the 2d Ising model on the square lattice~\cite{44-Onsager}.}  

Returning to the finite-temperature spin-spin Hamiltonian in section~\ref{secSpSp}, we note that spin-spin interactions fundamentally come from energetics coming from magnetic fields generated by the spinning charged particles. Putting the partition function in Eq.~\eqref{eqSpSp3} in this context, we see that it is given by
\begin{align}
Z(\beta,g,B) = e^{3 g^2 (m\beta)} + e^{-g^2 (m\beta)} (1+ e^{-g B (M \beta)} + e^{+g B (M \beta)}) ~, \label{eqYM2b4}
\end{align}
where $m$ and $M$ are mass-scales characteristic of the spin-spin and spin-field interactions, and $g$ is the QED gauge coupling constant. Now, this enlarged T-reflection operation sends $\beta \to -\beta$, $g \to -i g$ and $F_{\mu \nu} \to i F_{\mu \nu}$. Thus, the spin-spin partition function,
\begin{align}
Z(\beta,g,B) = Z(-\beta, -ig,i B)~. \label{eqYM2b5}
\end{align}
is completely invariant. 

There is further evidence that zero-temperature path integrals are invariant under reflecting $g_{\rm YM}^2 \to -g_{\rm YM}^2$. Vafa and Witten~\cite{35-VafaWitten} famously computed the path integral for topologically twisted ${\cal N} = 4$ SYM. These path integrals are written in terms of the complexified gauge-coupling $\tau := \t/2 \pi + i 4\pi/g_{\rm YM}^2$. The imaginary part of $\tau$ is $1/\alpha_{\rm YM}$. For a particular limit of ${\cal N} = 4$ SYM\footnote{Where the instanton moduli space is the compact $K3$-manifold.}, they found that the full path integral is equal to 
\begin{align}
Z(\tau) = \frac{1728}{E_4(\tau)^3 - E_6(\tau)^2}~\label{eqYM2b1}
\end{align}
where $E_4(\tau)$ and $E_6(\tau)$ are known as the classical Eisenstein series of weight four and six, and are defined by the following expression:
\begin{align}
E_{2k}(\tau):= \frac{1}{2 \zeta(2k)} \sum_{(m,n) \in \Z^2 \setminus (0,0)} \frac{1}{(m+n\tau)^{2k}} = E_{2k}(-\tau)~.~\label{eqYM2b2}
\end{align} 
It is manifest from the definition in Eq.~\eqref{eqYM2b2} that the Eisenstein series, and thus the Vafa-Witten path integral in Eq.~\eqref{eqYM2b1}, is invariant under reflecting $\tau \to -\tau$. Clearly, when $\theta = 0$ we have $\tau = i /\alpha_{\rm YM}$: these functions are invariant under $\alpha \to -\alpha$. This $\tau$-reflection invariance of these functions again offers evidence that supports the T-reflection prescription in section~\ref{secQMpert} applies to the perturbative corrections to gauge theory path integrals.

Before moving on, we should stress that this work of Vafa and Witten underlies an entire field devoted to studying Olive-Montonen duality~\cite{45-OliveMontonen} in gauge-theory path integrals. As emphasized above, the original Vafa-Witten partition functions are written in terms of modular forms which we expect to be invariant under T-reflection. However, more recent studies of Olive-Montonen duality have shown some of the exact path integrals are mock modular forms~\cite{46-VafaWitten2}. It is less clear that there is a nontrivial extension of mock modular forms that is well-behaved under T-reflection. We defer discussion of this important issue to Ref.~\cite{02T-rex2} and future work. 

In this subsection, we have given two examples where coupling-constant reflection is an exact invariance of a zero-temperature path integral for an interacting gauge-theory. This supports the previous analysis in subsection~\ref{secYM1a} that T-reflection invariance of perturbative expansions is qualitatively similar to T-reflection invariance of the anharmonic oscillator in quantum mechanics. 

Further, we have just provided an example that the necessary augmentation of T-reflection in the presence of interactions \emph{and} finite-temperatures cures the seeming non-invariance of the spin-spin interactions from section~\ref{secSpSp}. We now provide evidence in favor of this general picture, in the context of very intricate perturbative corrections to ${\cal N} = 4$ SYM at finite temperature.

\subsubsection{First corrections to the ${\cal N} = 4$ SYM partition function: Reflecting $\alpha$ and $\beta$}\label{secYM1c}

Ref.~\cite{04-QCD1} computed the zeroth-order terms in the path integral for Yang-Mills coupled to massless adjoint matter on $S^3_R \times S^1_{\beta}$. Following this, Spradlin and Volovich found the first pertubrative correction to the path integral for the special matter content corresponding to ${\cal N} = 4$ SYM~\cite{47-QCD9}. Explicitly, they found,
\begin{align}
\!\!\!\!\!\!
Z_{\rm MT}^{{\cal N} = 4}(q,\lambda_{\rm YM}) = Z^{(0)}_{\rm MT}(q) \bigg\{ 1 - \frac{\beta \lambda}{4 \pi^2} \sum_{k = 1}^{\infty}  \bigg(k \frac{\<D_2(q^k) \> }{1-z_v(q^k)} + \sum_{m = 1}^{\infty} \< PD_2(q^k,q^m) \>  \bigg) + {\cal O}(\lambda^2) \bigg\},\!\!\! \label{eqYM1b1}
\end{align}
where $\lambda := g_{\rm YM}^2 N$, $q = {\rm Exp}[-\beta/R]$, $z_v(q)$ is the vector partition function in Eq.~\eqref{eqYM0b}, $Z^{(0)}_{\rm MT}(q)$ is the path integral in Eq.~\eqref{eqYM0g}, and $\<D_2\>$ and $\< PD_2\>$ are the expectation values for the one-loop dilatation operator for the ${\cal N} = 4$ SYM conformal field theory. These have a detailed structure which will not concern us here.

Independent of their detailed structure, the $\< D_2\>$-terms are meromorphic functions in $q$ that converge at $q = 0$. As such, they have Taylor-series expansions about $q = 0$. Using this, we can show that the $\<D_2\>$-terms are invariant under the transformation in Eq.~\eqref{eqYM2b5}. The argument is startlingly independent of the details of the dilatation operator.

It is relatively straightforward to show this. First, note the string of equalities:
\begin{align}
F(q):= \sum_{n = 1}^{\infty} n f(q^n) 
&= \sum_{n = 1}^{\infty} n \bigg( \sum_{m=1}^{\infty} c_m q^{nm} \bigg) 
= \sum_{m=1}^{\infty} c_m \bigg(\sum_{n = 1}^{\infty} n  q^{nm} \bigg) \label{eqYM1b2} \\
&= \sum_{m=1}^{\infty} c_m \frac{q^{m}}{(1-q^m)^2} 
= \sum_{m=1}^{\infty} c_m \frac{(q^{-1})^{m}}{(1-(q^{-1})^{m})^2} = F(q^{-1}) ~. \label{eqYM1b3}
\end{align}
Second, recall that in the $k$-sum over $k \< D_2(q^k) \>/(1-z(q^k))$, the only $k$-dependence in the coefficients of $x^n$ come from the overall pre-factor of $k^{+1}$. Thus, we conclude
\begin{align}
\frac{\lambda}{4 \pi^2} \sum_{k=1}^{\infty} k\frac{\langle D_2(q^k) \rangle}{1-z(q^k)} = 
+ \frac{\lambda}{4 \pi^2} \sum_{k=1}^{\infty} k\frac{\langle D_2(1/q^k) \rangle}{1-z(1/q^k)} ~. \label{eqYM1b4}
\end{align}
The first class of terms in the first correction to the single-trace partition function in Ref.~\cite{47-QCD9} are invariant under T-reflection. Unfortunately, we have been unable to find an analogous argument that shows whether the $\< PD_2 \>$-terms are also invariant under $q \to 1/q$.

As emphasized in~\cite{47-QCD9} and follow-up papers, the $\< D_2\>$-terms dominate the perturbative correction to the path integral. Only these terms modify the location and residues of the Hagedorn poles. The $\< PD_2 \>$-terms are highly sub-dominant. Thus, the leading corrections to the ${\cal N} = 4$ SYM path integral are invariant if we send $\beta \to -\beta$ and $g_{\rm YM}^2 \to -g_{\rm YM}^2$:
\begin{align}
e^{i \pi/2}~ &Z^{(0)}_{\rm MT}(q) \bigg\{1 - \frac{\beta \lambda}{4 \pi^2} \sum_{k = 1}^{\infty} k  \frac{\<D_2(q^k) \> }{1-z_v(q^k)} + {\cal O}\big(\lambda \beta \< PD_2\>,\lambda^2\big)\bigg\} \nonumber\\
= ~& Z^{(0)}_{\rm MT}(1/q) \bigg\{1 - \frac{(-\beta)(- \lambda)}{4 \pi^2} \sum_{k = 1}^{\infty} k  \frac{\<D_2(q^{-k}) \> }{1-z_v(q^{-k})} + {\cal O}\big(\lambda \beta \< PD_2\>,\lambda^2\big)\bigg\} \label{eqYM1b5}
\end{align}
This invariance not only exactly matches the construction in this section and in section~\ref{secQMpert}, but it matches the exact pattern of T-reflection behavior of perturbative corrections to the anharmonic oscillator partition function in Eq.~\eqref{eqAHO5}.  

\subsection{Random Matrix Theory and truncations of thermal KK modes}\label{secRMT}

Our final example of T-reflection invariance in approximate descriptions of field theory path integrals concerns Ref.~\cite{48-QCD10}. In~\cite{48-QCD10}, they use a random matrix model to obtain analytic results for the universal portion of the QCD phase diagram. In this paper, Jackson and Vanderheyden truncate the full tower of Kaluza-Klein modes on the thermal circle to the just the first mode. Exploiting a mapping between the random matrix model fields and the order parameters for the QCD phase transition as a function of the chemical potential and temperature, they determine that the phase diagram is dictated by the following expression,
\begin{align}
Z(\mu,T) = \int d\sigma d\Delta {\rm Exp}[-4 N \Omega(\sigma,\Delta)]~,
\end{align}
where $\sigma \leftrightarrow \< \psi^{\dagger} \psi\>$ and $\Delta \leftrightarrow \< \psi \psi \>$, and 
\begin{align}
&\Omega(\sigma,\Delta) = A \Delta^2 + B \sigma^2 - \frac{1}{2} \bigg\{ (N_c-2) \big( \log[(\sigma + m -\mu)^2 + T^2] + \log[(\sigma + m -\mu)^2 + T^2] \big) \nonumber\\
&\qquad + 2 \big( \log[(\sigma + m -\mu)^2 + T^2 + |\Delta + \mu|^2] + \log[(\sigma + m -\mu)^2 + T^2+ |\Delta + \mu|^2] \big) \bigg\} .\!\!\!
\end{align}
This is clearly invariant under $T \to -T$. This invariance is inherited directly from the fact that path integrals integrate over all quantum fluctuations on all points on the thermal circle. The only remnant of the periodicity condition comes from the quantized KK momenta. As the KK mode energy is a simple rational function of quadratic in $T$, it is invariant under $T \to -T$.

\section{Differentiating T-reflection from possible candidates}\label{secNOT}

Ref.~\cite{01T-rex0} tabulated a series of example partition functions, path integrals, and supersymmetric indices that were invariant under reflecting temperatures to negative values. It is natural to try to ask if this T-reflection invariance, in all of its various guises and appearances, is due to one fundamental underlying physical explanation. Indeed, sections~\ref{sec0}--\ref{secYM} have built-up an argument that the experimental observation of T-reflection invariance of various path integrals and partition functions first observed in~\cite{01T-rex0} is due to a single underlying cause. We have proposed such a cause. 

Our explanation of T-reflection in this paper was constructed based on the two assumptions that (1) T-reflection has a common origin, and (2) that any unifying explanation for T-reflection had to match all of the known examples of what T-reflection can not be. 

Here is a list of physical principles that can not be related to T-reflection (with examples): 
\begin{itemize}
\item The negative temperatures involved in T-reflection are not obviously related to the normal ``negative temperatures'' encountered by spins in magnetic fields. In their normal occurrences, negative temperatures correspond to $[\frac{\partial S}{\partial E}] < 0$. In our context, we are merely studying the behavior of partition functions in the complex-$\beta$ plane.
\item T-reflection invariance is not a property of generic partition functions: only partition functions that are equal to QFT path integrals should have any ``nice'' behavior under T-reflection. For example, in section~\ref{secEnsembles} we showed that while the path integral for a theory of free photons is T-reflection invariant, its canonical partition function is not.
\item T-reflection is not at all obviously related to time-reversal invariance. Indeed, one can easily construct examples of quantum systems with non-zero external magnetic fields that are invariant under T-reflection. See appendix~\ref{secRing} for one such example.
\item T-reflection is not related to orientation reversal: orientation reversal is tied to complex conjugation. However, as T-reflection relates holomorphic Virasoro minimal model characters to (the same!) holomorphic minimal model characters, it cannot be tied to complex conjugation. (See appendix B of~\cite{02T-rex2} for a possible counterexample.) 
\item T-reflection invariance is not tied to conformal invariance: free massive QFTs are also invariant under T-reflection. The argument in section~\ref{secOld} and~\ref{secOld2} applies to free massless fields and to free massive fields equally.
\item Although it is easy to construct/find supersymmetric QFTs that are T-reflection invariant, T-reflection is not tied to supersymmetry: very, very few examples in this paper are supersymmetric.
\item T-reflection is not fundamentally tied to modular invariance, for the simple reason that it exists for non-conformal theories and for conformal theories outside of two-dimensions. That said, the arguments in section~\ref{secCircle} interplay with modularity in interesting ways. This interplay is a central focus of the following paper~\cite{02T-rex2}. 
\item T-reflection invariance is not tied to the strength of interactions between coupled oscillators. It is a property of consistent truncations, free systems, weakly interacting systems, and strongly interacting systems.
\item T-reflection invariance is not tied to exactly solvable quantum mechanical potentials nor of an exactly solvable quantum field theories. It is also a property of consistent truncations of QFTs with extremely rich perturbative and non-perturbative behavior, such as Yang-Mills, that are not known to be exactly solvable.
\end{itemize}
Recapitulating: we have argued in sections~\ref{secCircle} and~\ref{secQMpert} and~\ref{secYM} that T-reflection is a symmetry of path integrals for both free systems and their perturbative corrections. Further, as we have argued in section~\ref{secQMexact}, T-reflection may well be a property for exactly solved potentials in quantum mechanics, even though their explicit partition functions do not appear to be invariant under the most naive incarnation of T-reflection, where $\beta$ is sent to $-\beta$ while all other parameters are held fixed.

Having given this brief summary, we move on to the conclusions section, where we review the results derived in this paper and attempt to synthesize our current understanding of T-reflection. 

\section{Conclusions and further directions}\label{secEND}

T-reflection is an undeniable symmetry of the bosonic and fermionic harmonic oscillator partition functions: 
\begin{align}
Z_{\rm boson}(\beta) = \frac{1}{2 \sinh(\beta \omega/2)} \quad , \quad Z_{\rm fermion} = 2 \cosh(\beta \omega/2)~.
\end{align}
Because oscillators form the basis of our understanding of perturbative physics, it is reasonable to understand whether this surprising symmetry extends to general field theories.

As we have argued in section~\ref{sec0} and in Refs.~\cite{01T-rex0} and~\cite{02T-rex2}, it is also a manifest property of free QFTs that are composed of an infinite collection of free, decoupled modes. The argument in section~\ref{sec0} and in~\cite{01T-rex0} rests on a regularization that allows T-reflection invariance of the separate decoupled oscillators to survive in the field theory limit of these quantum systems. Further, in section~\ref{secCircle} of this paper we introduced a much more general argument for T-reflection invariance of general QFT path integrals, based on the lattice of points identified by circular compactification. This argument is extended to the two-torus in the following paper~\cite{02T-rex2}. 

Our new results and understanding come partitioned into three pieces: (1) our argument that T-reflection is a truly ubiquitous property of path integrals for QFTs at finite-temperature, (2) our understanding of T-reflection as an analytic continuation (which is especially important in understanding the interplay between T-reflection and perturbation theory),  and (3) a possible flaw in the above arguments. We describe these three pieces in sections~\ref{secENDu}, ~\ref{secENDs} and~\ref{secENDp}. We close with future directions in section~\ref{secENDf}.

\subsection{Arguments for the ubiquity of T-reflection}\label{secENDu}

First, in section~\ref{sec0} we explicitly showed free and interacting 2d CFTs are T-reflection invariant. This invariance holds at the level of individual characters for \emph{every} Virasoro minimal model. Thus, T-reflection invariance is decoupled from unitarity and parity (chirality), and holds for both free and interacting systems. 

Second, in section~\ref{secCircle} we argued that full QFT path integrals could be invariant under T-reflection. Our argument is solely based on how finite temperatures are input into QFT path integrals. Thus, it is independent of almost all details of the path integral. Finite temperatures enter path integrals by compactifying Euclidean time onto the thermal circle. 

After integrating over all fluctuations at all points on the (background) spacetime manifold, the only remnant of the thermal circle comes from quantized momenta along the thermal circle. These are in turn exclusively dictated by the lattice of points along the Euclidean time-direction that differ by integer multiples of the circumference, $\beta$. This lattice of identified points is equivalently generated by $+\beta$ and by $-\beta$. Thus, we expect the path integral to be invariant under reflecting $+\beta$ to $-\beta$. The remainder of the paper is devoted to probing this general argument. 

The main focus of~\cite{02T-rex2} is devoted to a similar claim in the two-dimensional generalization of this argument. In the context of 2d CFTs and modular forms, the relevant lattice is the two-dimensional lattice of points identified by compactifying the plane onto two circles $\C \to S^1_{1} \times S^1_{\tau}$ with circumference 1 and $\tau$. We call this lattice $\Lambda(\tau)$ and it is the collection of all complex numbers $\{ m + n\tau ~|~(m,n) \in \Z^2\}$. Up to re-scalings, this lattice is invariant under the discrete ``modular'' S-transformation $\tau \to -1/\tau$, ``modular'' T-transformation $\tau \to \tau+1$, and the (new) ``modular'' R-transformation $\tau \to -\tau$. These equivalences correspond to redundant encodings of the compact geometry of the two-torus. Path integrals for 2d CFTs on a two-torus are functions must be invariant under these redundancies. 

The argument given in section~\ref{secCircle} straightforwardly extends to 2d CFT path integrals, which are functions of this $\tau$-parameter. We find that they must be invariant under these so-called modular S- and T-transformations, as well as be invariant under the R-transformation. Crucially, this R-transformation is a direct analog of T-reflection~\cite{01T-rex0}. In the following paper~\cite{02T-rex2}, we study this mathematical claim in great detail.

Third, in section~\ref{secEnsembles} of this paper we emphasized that the logic suggesting \emph{path integrals} are invariant under T-reflections. This applies to partition functions only insofar as they are equal to path integrals. Because QFT path integrals incorporate all QFT processes, they necessarily include both single- and multi-particle processes. As such, we only expect full multi-particle partition functions to be invariant under T-reflection. In section~\ref{secEnsembles} we provide a characteristic example, that of the free Maxwell theory on $S^3 \times S^1$. Crucially, its single-particle partition function is \emph{not} invariant under T-reflection, while its multi-particle partition function is invariant under T-reflection.

The remainder of the paper is devoted to a more in-depth study of how one can realize T-reflection as an analytic continuation, and of how T-reflection interplays with perturbative expansions. Pursuing these questions suggests a possible synthesis of all of the available facts, which we now describe. 

\subsection{A(n attempted) sythesis}\label{secENDs}

The analysis of sections~\ref{secQMpert}, ~\ref{secQMexact} and~\ref{secYM} basically indicate that T-reflection is, in a very real sense, nearly vacuous. After all, to \emph{really} understand T-reflection invariance as an analytic continuation, we send $\beta H \to \beta(\theta) H(\theta) \to (-\beta) (-H)$. It is almost completely trivial that the path integral would be invariant under this transformation. However, as demonstrated in section~\ref{secSHO}, it is highly nontrivial to identify such a deformation. How does this analytic continuation fit into the evidence suggested in Ref.~\cite{01T-rex0}?

One answer could be that the notions of T-reflection as originally discussed in Ref.~\cite{01T-rex0} and this paper are completely different. In that original paper, we understood T-reflection largely as a discrete $\Z_2$ that sends $+\beta \to -\beta$. However, the predominant view in this paper and in~\cite{02T-rex2} is that T-reflection is most sensible in terms of a deformation of the path integral. At the end-points of the deformation, the Hamiltonian is Hermitian and the quantity $\beta H$ is positive and real. However, at any intermediate point along the deformation, this is not so. If there is a substantive difference between this paper and~\cite{01T-rex0}, then there is room for T-reflection to, for instance, not be related to a discrete redundancy in how the compact thermal-circle is encoded in the path integral. This could in principle rob T-reflection of its power to constrain vacuum energies. We think that this is, however, not the case.

Our perspective is that this analytic continuation is a way to realize the discrete T-reflection operation. The continuation between the two end-points is an \emph{analog} of the mapping-torus. Recall that mapping-tori are conventionally used as a way to realize discontinuous, large, coordinate- or gauge-transformations within a $d$-dimensional theory as coming from a smooth deformation of the theory when embedded within a $(d+1)$-dimensional space. Our analytic continuation is directly analogous: at either end of the deformation, the spacetime contour along which the Euclidean action is evaluated resides in the original one-dimensional space. However, at intermediate points along the path, the Euclidean action is a one-dimensional contour integral through a two-dimensional space. We emphasize that this perspective is speculative and supported only by the evidence in this paper and in the follow-up~\cite{02T-rex2}; it is logically possible that this picture could be incomplete or incorrect.

If the above interpretation is correct (and there are no counterexamples forcing us to modify our understanding), then T-reflection should be a discrete coordinate redundancy that is obtained by analytic continuation. In this context, it becomes clear how and why T-reflection has the power to constrain vacuum energies: Demanding a system be invariant under redundant descriptions of its spacetime manifold necessarily constrains its spacetime quantum numbers. As T-reflection concerns a redundancy in our descriptions of the (Wick-rotated and compactified) time-direction, it thus has the power to constrain vacuum energies.

This plausible explanation begs a follow-up question: If we demand that a QFT path integral be invariant under T-reflection, how can it be invariant \emph{up to an overall phase}? Our construction of T-reflection writes it as an analytic continuation of $\beta H \to \beta(\t) H(\t)$. At the beginning and end of the deformation, $H(0) \beta(0) = H(\pi) \beta(\pi)$, and $\beta = \beta(0) = -\beta(\pi)$. 

Thus, the Euclidean action is invariant under this continuous deformation:
\begin{align}
S_E = \beta H := S_E(0) \to S_E(\t) \to S_E(\pi) = S_E(0)~.
\end{align}
However, this does \emph{not} imply that the path integral will be invariant under T-reflection. The path integral, rather, is given by:
\begin{align}
\!\!
Z(e^{i0}\beta) := \int {\cal D}[x(0)] e^{-S_E(0)} 
\to Z(e^{i \t}\beta) = \int {\cal D}[x(\t)] e^{-S_E(\t)} 
\to Z(e^{i \pi}\beta) = \int {\cal D}[x(\pi)] e^{-S_E(\pi)} ~.
\end{align}
The analytic path that connects $+\beta$ to $-\beta$ is a symmetry of the Euclidean action: $S_E(+\beta) = S_E(-\beta)$. However, it is not necessarily a symmetry of the path integral \emph{measure}. As we explicitly saw in the simple harmonic oscillator, in section~\ref{secSHO}, the measure is \emph{not} invariant under this continuation: it picks-up the temperature-independent phase $e^{i \gamma}$.

\subsection{A possible gap and possible counterexamples}\label{secENDp}

To begin, we again retrace our understanding of T-reflection. In section~\ref{secCircle}, we related T-reflection invariance of a path integral to a redundancy in how the thermal circle is encoded in the path integral. Namely, finite temperatures enter into QFT by compactifying the temporal-direction of the spacetime manifold onto a compact Euclidean circle, $S^1_{\beta}$. Integrating over fluctuations on the circle leaves only the quantization condition for momenta along the thermal circle coming from its periodicity properties. These periodicity properties are identical for ``positive'' and ``negative'' temperatures. Hence, the path integral should not depend on the sign of the temperature: T-reflection invariance.

Further, in sections~\ref{secQMpert},~\ref{secQMexact}, and~\ref{secYM} we argued that this large $\beta$-reflection ``coordinate transformation'' can be realized as coming from the end-points of a complex rotation. This rotation sends $\beta$ to $e^{i \theta} \beta$ and then to $-\beta$. Now the Euclidean action, $S_E(\theta) = e^{i \t} \beta H(\t)$, matches at the two end-points of the rotation: $H(\pi) = -H(0)$ and thus $\beta H = (-\beta) H(\pi) = S_E(0)$. Thus, the path integral should be invariant under $\beta \to -\beta$, perhaps up to an ambiguity in defining the phase of the path integral measure.

In order for $H \to -H$ under $\beta \to -\beta$, we see that to preform T-reflection such that it is an invariance in a path integral for a nontrivial system, we must do more than a simple flip the sign of $\beta$ within the full QFT path integral. In order for $Z(\beta)$ to manifest T-reflection invariance, it also seems necessary to modify parameters in the action as $\beta \to e^{i \t} \beta \to -\beta$. Specifically, in our discussion of T-reflection invariance in sections~\ref{secQMpert} and~\ref{secYM}, we noted that in order for $H(\pi) = -H(0)$, we must preform a non-trivial action on the coupling constants that perturb the otherwise free and decoupled oscillators. 

The largest possible gap in this logic comes from the data provided by exactly solved quantum mechanical potentials, such as the Poschl-Teller potential. We now describe this possible gap. In all but one of our examples, we provided a fairly explicit prescription for how to show T-reflection invariance or non-invariance. The one exception is the Poschl-Teller potential in section~\ref{secPTV}. 

There we showed that the expansion of the exact partition function for small $\beta$,
\begin{align}
Z_{\rm Poschl-Teller}(\beta) 
&= \frac{1}{\sqrt{-2 \pi \beta E_1}} 
+ \sum_{\ell = 1}^{J}\bigg(\sum_{n = 0}^{\infty} \frac{C_n}{2^{n/2}} (-E_{\ell}\beta)^{\frac{n}{2}} \bigg)~,
\end{align}
is not invariant under flipping the sign of $\beta$. If we simply flip $+\beta$ to $-\beta$, then sequential terms in the infinite series expansion alternate in sign:
\begin{align}
\frac{1}{\sqrt{-2 \pi \beta E_1}} 
+ \sum_{\ell = 1}^{J}\bigg(\sum_{n = 0}^{\infty} \frac{C_n}{2^{n/2}} (-E_{\ell}\beta)^{\frac{n}{2}} \bigg)
\to
\frac{e^{i \pi/2}}{\sqrt{+2 \pi \beta E_1}} 
+ \sum_{\ell = 1}^{J}\bigg(\sum_{n = 0}^{\infty} \frac{C_n}{2^{n/2}} (+E_{\ell}\beta)^{\frac{n}{2}} \bigg)~. \label{eqPT2}
\end{align}
(Recall that $E_{\ell} = -\ell^2/2mL^2 < 0$.) Thus, this partition function seems to be non-invariant under reflecting $\beta \to -\beta$, without doing anything else to the potential. 

However, the point of section~\ref{secENDs} is to explain that if we merely reflect $\beta \to -\beta$ without doing any other operation on the Hamiltonian, then we would not expect the path integral to be invariant. Unpaired with any other operation this $\beta$-reflection is unlikely to yield the ``correct'' T-reflection. We argued that T-reflection should be understood as a rotation of $\beta \to e^{i \t} \beta$ and $H \to H(\t)$ such that the integration contour needed to define the path integral along this rotated trajectory is taken along the path of steepest ascent. We have shown that when this is done for the anharmonic oscillator, the Hamiltonian at the end of the deformation is equal and opposite to the original Hamiltonian: $H(\pi) = -H(0)$.

We have not done the analysis of steepest ascent for the Poschl-Teller potential. However, we conjecture that if this procedure were to be done, we would again have $H(\pi) = -H(0)$. If this were to happen, then we would have
\begin{align}
{\rm Poschl-Teller}~:~\beta E_{\ell} \to (e^{i \t} \beta) E_{\ell}(\t) \to (-\beta)(-E_{\ell})~, \label{eqPT4}
\end{align}
which could easily resolve the non-invariance of the small-$\beta$ expansion in the Poschl-Teller partition function in Eq.~\eqref{eqPT2}. We leave this analysis to future work.

The above analysis can also plausibly show that the Poschl-Teller potential from section~\ref{secPTV} is a counterexample to the general argument in section~\ref{secCircle} for ubiquitous T-reflection invariance. The reason for this is as follows. In order to take $H \to -H$, we must find an operation that reverses the sign of the kinetic term. However, as we can see from the harmonic oscillator in section~\ref{secQMpert} (and perturbative QFT in section~\ref{secYM}):
\begin{align}
x \to i x \implies H_{\rm Kin} = (i \d_x)^2 = - (\d_x)^2 \to -H_{\rm Kin} = -(i \d_x) = + (\d_x)^2 \label{eqPT5}~.
\end{align}
However, sending $x \to i x$ qualitatively changes the structure of the Poschl-Teller potential:
\begin{align}
{\rm Poschl-Teller}~:~V(x) = -J(J+1){\rm sech}^2(x) \to V(i x) = -J(J+1) \sec^2(x) \label{eqPT6}~.
\end{align}
It is difficult to find field transformations that are consistent with Eqs.~\eqref{eqPT5} and~\eqref{eqPT6} that also send $H \to H(\t) \to H(\pi) = -H$. 

At this point it is not clear whether such field transformations are necessary to show T-reflection invariance. Indeed, to show T-reflection invariance in the context of deforming the path integral as a function of $e^{i \t} \beta$, we merely need to find the contour of steepest ascent as $\theta \to \pi$. We did not need to define a field transformation. Of course, the existence of such a field transformation almost automatically implies T-reflection invariance. Indeed, this is how we understood the perturbative corrections to the anharmonic oscillator and to free gauge fields in sections~\ref{secQMpert} and~\ref{secYM}. 

Absent a proof that contours that extremize a deformed Euclidean action and explicit Hamiltonian deformations which also extremize the Euclidean action are necessarily equivalent, it is quite possible that the tension between Eq.~\eqref{eqPT5} and~\eqref{eqPT6} is a false signal. It is quite possible that the Poschl-Teller potential is indeed invariant.

\subsection{Future directions}\label{secENDf}

\subsubsection{Pursuing inconsistencies and possible counterexamples}

Here, we emphasize possible inconsistencies in the general picture developed in this paper, as well as possible counterexamples. To begin, it is important to emphasize that there may be a clash between the general pictures put forth in sections~\ref{secCircle} and~\ref{secQMpert}. Our argument that T-reflection is a redundancy in how the thermal circle is encoded in the path integral is very general. Furthermore, it naturally extends to a broader picture in the context of modular forms and 2d CFTs, as discussed in~\cite{02T-rex2}. In this picture, T-reflection is much more of a discrete $\Z_2$ operation. However, this simple picture seems to clash with complexity of anharmonic oscillators as $\beta \to e^{i \t} \beta \to -\beta$ discussed in section~\ref{secQMpert}. It will be very important to resolve whether the mismatch of complexity of continuations vs. redundancies of identified points implies that discrete T-reflection as discussed in section~\ref{secCircle} and~\cite{02T-rex2} is distinct from our analytic continuation in sections~\ref{secQMpert} and~\ref{secYM}. 

Regardless, in this paper we developed a framework for understanding T-reflection as an analytic continuation. This analytic continuation does not actively ``invert'' the spectrum of a Hamiltonian in the most obvious way that T-reflection might act. 
Our general, new, framework to analyze T-reflection (non)invariance of approximate and exact systems extends far beyond the original analysis of~\cite{01T-rex0}.

Presently, we do not know if the Poschl-Teller potential in section~\ref{secPTV} is a counterexample to the general claim in this paper that QFT path integrals are T-reflection invariant. An analysis similar to the transformations in section~\ref{secSHO} for the Poschl-Teller potential poses an important test of the narrative in this paper. Here, T-reflection invariance derives from a deformation of the path integral contour that smoothly connects $\beta$ to $-\beta$ via $S_E(\pi) = S_E(0)$. 

Similarly, one could search for a similar deformation in the simpler case of an attractive $\delta$-function potential. This potential is also exactly solvable, and is considerably simpler: it has exactly one bound-state, and a continuum that is almost identical to the continuum for the free particle. The presence of \emph{both} bound-states and continuum-states distinguishes these solvable potentials from the other examples in this paper, where we consider systems whose excitation spectrum is either entirely discrete (a confining potential) or entirely a continuum. At present, it is logically possible that T-reflection is a symmetry of potentials with spectra that are either entirely discrete or entirely continuous.

If no such deformation exists for these exactly solved potentials, then we might reasonably conclude that the Poschl-Teller potential is a counterexample to the general argument in section~\ref{secCircle}. This is because quantum mechanics is a QFT in $(0+1)$-dimensions. Consequently, the the partition functions for these potentials \emph{should} constitute path integrals for valid quantum field theories in one-dimension.  Any non-invariance would lead to tension. 

There are two obvious places where non-invariance could be explained. First, there could be hidden subtleties in our naive equation of statistical mechanical partition functions for QM and finite temperature path integrals in $(0+1)$-dimensional QFT. Second, non-invariance can imply hidden assumptions in the argument in section~\ref{secCircle} that all QFT path integrals are T-reflection invariant. The second option seems more likely. Presently, though, we do not regard any of the extant path integrals as counterexamples for T-reflection. 

\subsubsection{Past uses of T-reflection}

We reemphasize that T-reflection has already had concrete and highly non-trivial applications in the literature. Thus, independent of its ultimate fate as a genuine property of all finite-temperature field theory path integrals, it has already proven useful. It was first discovered in the context of $SUN(N)$ Yang-Mills on $S^3 \times S^1$~\cite{04-QCD1} at large-$N$, coupled to arbitrary amounts of scalars and fermions which transform in the adjoint representation of $SU(N)$. With this observation, we were able to immediately find three highly non-trivial aspect of this theories: (1) $(-1)^F$-twisted \emph{non-supersymmetric} Yang-Mills coupled to adjoint matter has a hidden symmetry between bosons and fermions~\cite{05-QCD2} unrelated to supersymmetry, (2) that this theory has an identically vanishing vacuum energy~\cite{06-QCD3, 07-QCD4} that holds for arbitrary amounts of scalars and fermions, and (3) that this theory is inherently two-dimensional in nature~\cite{41-QCD6, 40-QCD5}. It has also been used to understand properties of higher-spin theories~\cite{49-tref1}, the 3d-3d correspondence~\cite{50-tref2}, the Yang-Baxter equation~\cite{51-tref3}, and topological entanglement entropy of large-$N$ gauge theories that exhibit Hagedorn phase transitions~\cite{52-tref4}.

\subsubsection{Forthcoming publications}

There are three upcoming publications that are directly related to T-reflections. First, in Ref.~\cite{02T-rex2} we directly extend the analysis in sections,~\ref{sec0},~\ref{secCircle} and~\ref{secSHO} to two-dimensional conformal field theories on the compact two-torus. In the process, we argue that all rational and algebraic functions of the classical Eisenstein series are invariant under reflecting $\tau \to -\tau$. We point-out that the original definition of the Eisenstein series $E_{2k}(\tau)$ is invariant under reflecting the temperature parameter to negative values by definition: $E_{2k}(\tau) = E_{2k}(-\tau)$. From this, we argue that all 2d CFT torus path-integrals are invariant under T-reflection.

Second, in the forthcoming Ref.~\cite{03-Lfunctions} we pursue further mathematical corollaries of T-reflection. Specifically, we study whether an analog of the $q$-inversion analysis championed in section~\ref{sec0} extends to a much broader class of modular forms that have infinite product expansions. Specifically, in~\cite{53-BP1, 54-BP2} an infinite sequence of modular forms were shown to have infinite product expansions of the form $q^{-F(0)} \prod (1-q^n)^{F(n)}$, and thus resembles an infinite collection of decoupled harmonic oscillators. Demanding T-reflection invariance then seems to require the following sum-rules:
\begin{align}
H(\tau) = q^{-F(0)} \prod_{n =1 }^{\infty} (1-q^n)^{F(n)} \implies 
\left\{ 
\begin{matrix} 
e^{i \gamma} = (-1)^{-\sum_{n \geq 1} F(n)~n^0} = (-1)^k \\
q^{-F(0)} = q^{-\sum_{n \geq 1} F(n)~n^1 /2} \quad \qquad~
\end{matrix}
\right. \label{eqBZsumG}
\end{align}
These identities resemble special values of a zeta-function (L-function) attached to the modular form $H(\tau)$, evaluated at $s = -1$ and at $s = 0$. However all of these products save a set of measure zero have $F(n)$ that grow exponentially: $F(n) \sim e^{\beta_F n}$ (Hagedorn growth). This growth overwhelms the $n^{-s}$ suppression from the Dirichlet-series definition of zeta-functions. 

To evaluate these sums in~\cite{03-Lfunctions} we make a crucial and natural generalization of modular L-functions: We define L-functions for \emph{meromorphic} modular forms that have poles. Crucially, poles in a modular form imply that its $q$-series coefficients grow exponentially. We show that exponential growth of the $F(n)$ comes exactly from poles. We then define the associated zeta-functions/L-functions, and use these L-functions to verify the sum-rules in Eq.~\eqref{eqBZsumG}.

Third, in the forthcoming Ref.~\cite{08-SPT-R} we interpret to use the T-reflection phase $Z(-\beta) = e^{i \gamma} Z(\beta)$ to classify symmetry protected topological phases (SPT phases) in condensed matter systems. As emphasized in~\cite{19-SPT1} and~\cite{20-SPT2}, if a path integral for gapless excitations on the boundary of an insulator acquire non-trivial phases under redundant relabelings of its underlying spacetime manifold, then the insulator has SPT-topological order. As emphasized in section~\ref{secCircle}, T-reflection invariance can be thought of as a redundant way to encode the thermal circle in the path integral. As such, T-reflection may be useful in classifying SPT-order if the boundary CFT for a condensed-matter insulator has non-trivial T-reflection phases.

\acknowledgements 

This research was supported by a JSPS Visiting Postdoctoral Fellowship at the Kavli Institute for Physics and Mathematics of the Universe (IPMU), the Niels Bohr International Academy (NBIA), and from a Carlsberg Distinguished Postdoctoral Fellowship at the NBIA. We thank Aleksey Cherman and Masahito Yamazaki for collaboration during early stages of this project. Further, we thank Aleksey Cherman for detailed comments on the draft and crucial conversations on Lefshetz thimbles and T-reflection for (an)harmonic oscillators. We thank Aleksey Cherman, Simon Caron-Huot, Matthias Wilhelm and Marcus Spradlin for discussions regarding T-reflection for perturbative corrections to the path integral for gauge-theories in four-dimensions. We thank Andrew Jackson for his comments regarding T-reflection and random matrix theory. Finally, we thank John Duncan, Shunji Matsuura, and Shinsei Ryu for many helpful conversations and for collaboration on related projects~\cite{03-Lfunctions, 08-SPT-R}.

\appendix

\section{T-reflection in diverse quantum mechanical systems}\label{secRandom}

It is very simple to find quantum mechanical systems that are, and are not, invariant under T-reflection. Here we present three interesting systems in non-relativistic quantum mechanics: the infinite square-well, the particle on a ring, and the Hydrogen atom. 

In sections~\ref{secWell} and~\ref{secRing} we show that the partition function for both (a) the infinite square-well and (b) the particle on a ring are given by modular forms that are proportional to the partition functions for 2d CFTs discussed in section~\ref{sec0}. This persists even when a magnetic field is threaded through the ring. This is interesting, as magnetic fields break time-reversal invariance. Sending $L$, equivalently the radius of the ring or the width of the square potential, to infinity we recover the standard partition function for a free non-relativistic gas of particles. It is also invariant under T-reflection. In section~\ref{secHydrogen}, we study the partition function for the exactly solved bound states of Hydrogen; it diverges at $T \neq 0$.

\subsection{The infinite square-well}\label{secWell}

The non-relativistic Schrodinger equation in the infinite square-well has a unique ground-state, the constant function, and a pair of states with momentum $k = \pm 2 \pi/L$, where $L$ is the width of the well. Thus, its partition function is,
\begin{align}
Z(\beta) = 
1 + \sum_{k = 1}^{\infty} 2 e^{-\beta \tfrac{k^2}{2mL^2}} = 
\sum_{k \in \Z} q^{k^2/2} = 
\prod_{n = 1}^{\infty}(1-q^n)(1+q^{n-1/2})^2~,
\end{align}
where we have defined the dimensionless variable $q:= {\rm Exp}[-\beta/(mL^2)]$. 

The methods of section~\ref{sec0} imply that this infinite product is invariant under $q \to 1/q$, modulo the phase $e^{i \gamma} = (-1)^{\zeta(0)}$. Thus, the infinite square-well is T-reflection invariant.

\subsection{Particle on a ring}\label{secRing}

First, we consider the particle on a ring with zero magnetic flux. In this case, there is a constant solution: the zero-mode. Further, there are two modes for each periodic solution: one with counter-clockwise angular momentum, the other with clockwise angular momentum. If we call the circumference of the ring $L$, then at zero-flux, its spectrum is identical to that of the particle in the box:
\begin{align}
Z(\beta, \phi = 0) = 
\sum_{k \in \Z} e^{-\beta \tfrac{(k+0)^2}{2mL^2}} = 
\prod_{n = 1}^{\infty}(1-q^n)(1+q^{n-1/2})^2~.
\end{align}
It is also invariant under T-reflection up to a phase of $e^{i \gamma} = (-1)^{\zeta(0)}$. 

Second, if we thread a magnetic flux $\phi$ through the ring, this shifts the angular momentum $k/L \to (k+ \phi)/L $. The resultant partition function is,
\begin{align}
Z(\beta, \phi) = 
\sum_{k \in \Z} e^{-\beta \tfrac{(k+\phi)^2}{2mL^2}} = 
q^{\phi^2/2}\prod_{n = 1}^{\infty}(1-q^n)(1+q^{n-1/2+\phi}) (1+q^{n-1/2-\phi})~.
\end{align}
This is also invariant under T-reflection up to a phase of $e^{i \gamma} = (-1)^{\zeta(0)}$, which can be verified by simple analysis of the Hurwitz zeta-function directly analogous to those in Eq.~\eqref{prodChar}.

Note that these two partition functions are modular. This should not be overly surprising, as these systems are effectively defined on the two-torus, $S^1_L \times S^1_{\beta}$. Further, note that the T-reflection phase $(-1)^{-1/2}$ of each of these ``free'' non-relativistic systems is exactly reproduced by the continuum limit $L \to \infty$. At $L \to \infty$, the free system is simply a non-relativistic gas, with partition function
\begin{align}
Z(\beta) = \frac{L}{\lambda_{\rm tdB}} = L \sqrt{\frac{m}{2 \pi \beta}}~,
\end{align}
where $\lambda_{\rm tdB}$ is the thermal de-Broglie wavelength of the particle.  Importantly, this free system has the same T-reflection in all three forms: $(-1)^{\zeta(0)}=e^{i \gamma}$. Coherence between these distinct T-reflection phases gives added strength and confidence that the various physical pictures for T-reflection given in this paper agree with each other and are consistent.

\subsection{Hydrogen}\label{secHydrogen}

This last example is actually a non-example. But it is in the natural trajectory of both this appendix and sections~\ref{secQMpert} and~\ref{secQMexact}. Namely: In all this discussion of T-reflection for exactly solved systems, one may have wondered about the partition function of bound states Hydrogen and the $1/r$ Coulomb-potential. 

This would be misguided, for an interesting reason: The exactly solved bound states for the $1/r$ potential in three-dimensional quantum mechanics accumulate at the threshold for the continuum. Here, the continuum represents ionized Hydrogen. Hydrogen's dynamics are not fully captured by its bound states; ionized states are also very important. 

One way to see this is to consider the partition function truncated to bound states:
\begin{align}
Z_{E<0}(\beta) := \sum_{n = 1}^{\infty} d_n e^{-\beta E_n} \quad , \quad d_n = n^2 \quad {\rm and} \quad E_n = \Delta \left(1- \frac{1}{n^2} \right)~.
\end{align}
Here $d_n$ and $E_n$ are the well-known degeneracies and energies of the attractive $1/r$ potential. Because $1-1/n^2 \to 0$ as $n \to \infty$, we have an accumulation point at $E = 0$. At any non-zero temperature $T \neq 0$, $Z_{E<0}(\beta)$ diverges: the ionized continuum states must be included. 

\section{T-reflection for correlation functions in the Ising minimal model}

In section~\ref{sec0}, we reiterated the claim that the path integrals for 2d CFT minimal models are invariant under T-reflection. It is natural to ask: If the path integral is invariant under T-reflection, what about correlation functions? Are correlation functions invariant under T-reflection? In this section we give examples of correlation functions, in the 2d CFT Ising minimal model and in certain chiral (holomorphic) 2d CFTs, that are demonstrably invariant under T-reflection. Because this discussion is somewhat outside the narrative of this paper, we have relegated it to an appendix. Our notation here is that of chapters 8, 10, and 12 of Ref.~\cite{21-Yellow}. 

A standard result is that two-point correlation functions between fermions on $T^2$ in the CFT describing the 2d Ising model on the square lattice, i.e. the $\mathcal{M}(3,4)$ 2d CFT, is
\begin{align}
\langle \psi(z,\tau) \psi(w,\tau) \rangle_{a} = \frac{\theta_a(z-w,\tau) \partial_x \theta_1(0,\tau)}{\theta_a(0,\tau) \theta_1(z-w,\tau)} \label{Ising2pt1}
\end{align}
where $\partial_x \theta(0,\tau) = \partial_x \theta(x,\tau)|_{x\rightarrow0}$ and $a \in \{ 2, 3, 4\}$. We would like to see if 
\begin{align}
\langle \psi(z,\tau) \psi(w,\tau) \rangle_{a} = e^{i \phi} \langle \psi(z,-\tau) \psi(w,-\tau) \rangle_{a} \label{Ising2pt2}
\end{align} 
for some fixed $\phi$ which does not depend on any dynamical variables such as position or $\tau$ itself. We can see that this equation holds, with $\phi = \pi$, directly. First, note that all factors in Eq.~\eqref{Ising2pt1} save for $\partial_x \theta_1(0,\tau)$ are T-reflection invariant by the arguments in the preceding sections. So, now, we would like to check if $\partial_x \theta_1(0,\tau)$ has nice properties under T-reflection. To do this, we write $\theta_1(z,\tau)$ in its infinite product form,
\begin{align}
\theta_1(z,\tau)
&= 2 q^{1/8} \sin (\pi z) \prod_{n=1}^\infty (1 - q^n) (1 - q^n e^{2\pi i z})(1 - q^n / e^{2\pi i z})
\end{align}
Now, hitting this with $\partial_z$ yields, after some manipulation,
\begin{align}
\partial_z \theta_1(z,\tau) &= \left\{2 \pi \cos(\pi z) + 8 \pi \sin(\pi z) \sin(2 \pi ) \sum_{n = 1}^{\infty} \frac{q^n}{1-2 \cos(2\pi z) q^n + q^{2n}} \right\} \nonumber\\ 
&\times \left( q^{1/8} \prod_{n = 1}^{\infty} (1-q^n)(1 - q^n e^{2\pi i z})(1 - q^n / e^{2\pi i z}) \right) \, .
\end{align}
Note that each block in the derivative is invariant under T-reflections. Further, note that in the limit that $z \to 0$, we have
\begin{align}
\partial_z \theta_1(z,\tau)\bigg|_{z \to 0} &= 2 \pi \eta(\tau)^3 \, . \label{DZ0}
\end{align}
As we can clearly see, the two-point correlation function in Eq.~\eqref{Ising2pt1} can be written entirely in terms of T-reflection covariant building-blocks. Because the correlation function~\eqref{Ising2pt1} has modular weight $w$, its T-reflection phase is $(-1)^{w+1}$ (as the $z$-derivative shifts $w \to w+1$).

We expect that as long as the correlation functions in 2d CFTs are modular forms (or seqsuilinear combinations of rational functions of modular forms and Jacobi forms), they should be invariant under T-reflection. The reason for this is that we expect modular forms and Jacobi forms to be invariant under T-reflection, up to an overall phase from their modular weight (and possible action on the fugacities of Jacobi forms). The grounds for this expectation are the main focus of the follow-up work~\cite{02T-rex2}. Finally, we expect general correlators to also be invariant, modulo T-reflection phases.

\end{document}